\documentclass[pdflatex,sn-mathphys-num, iicol]{sn-jnl}

\usepackage{graphicx}%
\usepackage{multirow}%
\usepackage{amsmath,amssymb,amsfonts}%
\usepackage{amsthm}%
\usepackage{mathrsfs}%
\usepackage[title]{appendix}%
\usepackage{xcolor}%
\usepackage{textcomp}%
\usepackage{manyfoot}%
\usepackage{booktabs}%
\usepackage{algorithm}%
\usepackage{algorithmicx}%
\usepackage{algpseudocode}%
\usepackage{listings}%
\usepackage[normalem]{ulem}%

\theoremstyle{thmstyleone}%
\theoremstyle{thmstyletwo}%

\theoremstyle{thmstylethree}%

\usepackage{lineno}
\usepackage{bm}
\usepackage{placeins}
\usepackage{float}
\usepackage{array}
\usepackage{color} 
\usepackage{tikz-feynman}
\tikzfeynmanset{warn luatex=false}
\usepackage{caption} 

\newcommand{\eg}{\textit{e.g.}}

\newcommand{\ie}{\textit{i.e.}}
\newcommand{\dd}{\mathrm{d}}

\newcommand{\raa}{R_{\rm AA}}
\newcommand{\npart}{N_{\rm part}}

\newcommand{\pT}{p_{\rm T}}

\definecolor{mypink1}{RGB}{235, 82, 247}

\begin{document}

\title[Coupled charm and charmonium transport in a strongly coupled quark-gluon plasma]{\boldmath Coupled charm and charmonium transport in a strongly coupled quark-gluon plasma}


\author*[1]{\fnm{Kaiyu} \sur{Fu}}\email{kaiyufu@tamu.edu}

\author[1,2]{\fnm{Biaogang} \sur{Wu}}\email{bwu8@kent.edu}

\author[1]{\fnm{Ralf} \sur{Rapp}}\email{rapp@comp.tamu.edu}

\affil*[1]{\orgdiv{Cyclotron Institute and Department of Physics and Astronomy}, \orgname{Texas A\&M University}, \orgaddress{\city{College Station}, \state{TX}, \postcode{77843-3366}, \country{USA}}}

\affil[2]{\orgdiv{Department of Physics},
\orgname{Kent State University},
\orgaddress{\city{Kent}, \state{OH},
\postcode{44242}, \country{USA}}}


\abstract{
The quark--gluon plasma (QGP) is a strongly coupled medium in which both open and hidden charm particles experience substantial nonperturbative interactions.
This poses a major challenge for a quantitative description of charmonium transport in ultra-relativistic heavy-ion collisions, as it requires a mutually consistent treatment of pertinent transport coefficients.
In this work, we present a coupled charm-charmonium transport framework for a strongly coupled QGP based on thermodynamic $T$-matrix interactions with recent constraints from Wilson-line correlators (WLCs) computed in lattice QCD. For the first time, the same underlying 
heavy-light interactions and in-medium spectral functions are used to self-consistently evaluate charm-quark diffusion and charmonium kinetics. 
In particular, the charmonium
equilibrium limit, a critical transport parameter for regeneration, is evaluated in the presence of broad spectral functions.
Charm-quark diffusion is simulated via Langevin dynamics and coupled to a
Boltzmann equation for charmonium dissociation and regeneration.
The equilibrium limit of the statistical model is recovered once charm quarks thermalize, and its extension to describe off-equilibrium is constructed.
Preliminary applications to charmonium observables in Pb--Pb collisions at the LHC capture the measured centrality and momentum dependence fairly well. 
}




\maketitle

\tableofcontents

\section{Introduction} 
\label{sec:intro}
The production of charmonium in ultra-relativistic heavy-ion collisions (URHICs) has long been recognized as a suitable probe of the medium modifications of the fundamental Quantum Chromodynamics (QCD) force in the quark–gluon plasma (QGP). Since the original idea that color screening in a deconfined medium suppresses the formation of charmonium bound states~\cite{Matsui:1986dk}, extensive theoretical and experimental investigations have revealed that the observed yields result from an interplay of suppression and regeneration mechanisms in the fireball of URHICs~\cite{Rapp:2008tf,Braun-Munzinger:2009dzl,Mocsy:2013syh, Andronic:2024oxz}.
In particular, abundant production of $c\bar{c}$ pairs in central collisions of Pb nuclei at the Large Hadron Collider (LHC) 
led to the prediction of substantial charmonium regeneration~\cite{Braun-Munzinger:2000csl,Thews:2000rj,Grandchamp:2001pf}, which was confirmed by experiment.

Quantitative descriptions of charmonium production have been developed using dynamical transport approaches that evaluate the time evolution of the bound states in the expanding fireball. The inelastic reaction rates driving dissociation and regeneration are central to this evolution, as they encode the medium-modified binding energies and radii, as well as the spectral properties of the charmonia which quantify their coupling to the thermal partons of the QGP. Many existing transport models rely on perturbative heavy–light interactions and usually require phenomenological $K$-factors to reproduce the suppression of excited states, such as $\psi(2S)$, in URHICs~\cite{Du:2015wha,Wu:2024gil}.
These descriptions do not fully capture the dynamics of the strongly coupled QGP (sQGP)~\cite{Shuryak:2004cy,vanHees:2005wb}, as evidenced by the small heavy-quark (HQ) diffusion coefficients extracted from phenomenology~\cite{Rapp:2018qla} and lattice QCD~\cite{Altenkort:2023eav}, as well as by lattice-QCD based determinations of the HQ potential that exhibit pronounced nonperturbative features~\cite{Tang:2023tkm}.

More recent developments in quarkonium transport have moved beyond the perturbative regime through the use of the thermodynamic $T$-matrix approach~\cite{Wu:2025lcj}.
This framework incorporates nonperturbative heavy–light scattering, self-consistent in-medium parton spectral functions (which are strongly broadened), and medium-modified heavy-quark potentials constrained by lattice-QCD calculations of Wilson-line correlators (WLCs)~\cite{Bala:2021fkm}. When deploying pertinent transport coefficients for HQ diffusion to open heavy-flavor (HF) phenomenology at the LHC, a decent explanation of HF hadron data emerges~\cite{Krishna:2025bll}.
Quarkonium reaction rates have been computed from the same underlying interaction as that governing open HF transport~\cite{Wu:2025hlf}, thus encapsulating the strongly coupled nature of the QGP. Embedding these rates into quarkonium transport models thus establishes a direct connection between quarkonium observables and the microscopic interaction strength of the sQGP, which has been carried out for bottomonia in Pb--Pb and p--Pb collisions at the LHC in Refs.~\cite{Wu:2025lcj,Thapa:2025jua}.

Strong interactions among the partons in the QGP give rise to broad in-medium spectral functions.
The widths of these spectral functions can enhance quarkonium dissociation rates by enabling off-shell scattering which increases the available phase-space for the outgoing light partons and charm quarks.
However, a recent $T$-matrix study in the complex-energy plane revealed that large in-medium widths do not necessarily imply the disappearance of quarkonium states~\cite{Tang:2025ypa}.
Instead, quarkonium poles can persist to high temperatures while acquiring large widths, indicating robust $Q\bar Q$ correlations in the form of resonances over a broad temperature range.
This extends the effective regeneration window during the fireball evolution and, in combination with large reaction rates, drives quarkonia toward their equilibrium limits rather rapidly.
A proper treatment of the charmonium equilibrium limit in the presence of broad spectral functions is therefore essential for reliably determining the final charmonium yields.

Another essential ingredient for the charmonium equilibrium limit is the kinetic evolution of charm quarks.
Charm and anticharm quarks produced in initial hard scatterings are far from thermal equilibrium but gradually approach it through diffusion and drag in the medium.
Their time-dependent momentum distributions directly impact the magnitude and momentum dependence of regenerated charmonia~\cite{Grandchamp:2002wp,Song:2012at}.
Langevin simulations provide an effective framework for describing HQ diffusion in the QGP~\cite{Moore:2004tg,vanHees:2004gq,Mustafa:2004dr} and enable one to track the time evolution of charm quarks throughout the evolution of the fireball~\cite{Du:2022uvj}.
The key inputs to these simulations are the HQ transport coefficients, which are, in principle, closely related to the quarkonium reaction rates.

In the present paper we build on earlier work~\cite{Du:2022uvj} but with several key improvements, that, in particular, are based on much more realistic interactions rooted in the sQGP, along with the required technical developments. 
While Ref.~\cite{Du:2022uvj} relied on perturbative reaction rates, we here, for the first time, employ a universal underlying heavy-light
interaction together with off-shell spectral functions to compute both charm-quark and charmonium transport coefficients.
Time-dependent off-equilibrium charm-quark distributions obtained from Langevin simulations are coupled to charmonium regeneration in the Boltzmann transport equation governing charmonium kinetics. In particular, the equilibrium limit is formulated (and validated) through the ratio of dissociation and regeneration rates, which enables
a controlled implementation of the in-medium
spectral functions of charm quarks and thermal partons.
As a preliminary test, we also compute charmonium observables at the LHC using a schematic fireball expansion for the medium evolution (albeit with an improved equation of state compatible with lattice QCD).

The remainder of this paper is organized as follows.
In Sec.~\ref{sec:langevin} we briefly recall the Langevin dynamics used to simulate charm-quark diffusion in the QGP and the calculation of the corresponding transport coefficients from the in-medium $T$-matrix.
In Sec.~\ref{sec:boltzmann}, we introduce the Boltzmann equation for charmonium transport including the derivation of charmonium dissociation and regeneration rates from the same $T$-matrix interactions.
In particular, we obtain the regeneration rates using
off-shell spectral functions while satisfying detailed balance and ensuring the correct equilibrium limit, stipulating the connection between off-equilibrium charm-quark and charmonium transport.
In Sec.~\ref{sec:lhc}, we turn to our applications to observables, specifying the phenomenological inputs from $pp$ spectra, cold-nuclear-matter effects, and the time evolution of the medium and presenting our results for charmonium nuclear modification factors ($\raa$) in Pb--Pb collisions at the LHC.
Finally, Sec.~\ref{sec:conclusions} contains a summary and outlook.

\section{Heavy-quark transport}
\label{sec:langevin}
In this section, we recall the relativistic Langevin approach to
simulate HQ diffusion in the QGP (Sec.~\ref{ssec:lang})
and discuss the calculation of the charm-quark transport
coefficient, \ie,  the relaxation rate, from the nonperturbative
$T$-matrix approach (Sec.~\ref{ssec:A}).

\subsection{Langevin simulation}
\label{ssec:lang}
%
Heavy quarks, \ie, charm or bottom, have masses, $m_Q$, that are much larger than typical medium temperatures in URHICs ($m_Q \gg T$) and momentum transfers in a single scattering, which are small compared to the momentum ($\Delta p \ll m_Q$). Under these conditions, the evolution of the charm-quark phase-space distribution, $f_c$, can be described by a Fokker–Planck equation~\cite{vanHees:2004gq},
\begin{equation}
    \begin{aligned}
&\frac{\partial}{\partial t} f_c(t,\mathbf{p})=\\
& \frac{\partial}{\partial p_i}
\left[
\Gamma(\mathbf{p})\, p_i \, f_c(t,\mathbf{p})
+
D(\mathbf{p})\, \frac{\partial}{\partial p_i} f_c(t,\mathbf{p})
\right]\,,
\end{aligned}
\label{eq:FP}
\end{equation}
where $\mathbf{p}$ is the charm-quark three-momentum
and $\Gamma(\mathbf{p})$ and $D(\mathbf{p})$ are the friction and momentum diffusion coefficients, respectively.

Relativistic Langevin dynamics in expanding media has been widely employed in phenomenological studies of HF transport
\cite{Svetitsky:1987gq,Moore:2004tg,Rapp:2009my,He:2013zua}.
It preserves the correct local thermalization
behavior dictated by the underlying Fokker–Planck equation, while enabling an efficient simulation of the diffusion process
in space-time–dependent backgrounds.

In the post-point scheme~\cite{He:2013zua},
the equilibrium conditions (Einstein relations) read
\begin{equation}
\begin{aligned}
D(\mathbf{p}) =& \Gamma(\mathbf{p})\, E\, T\,,\\
\Gamma(\mathbf{p}) =& A(\mathbf{p}) + \frac{1}{E} \frac{\partial D}{\partial E}\,,
\end{aligned}
\end{equation} 
where $A(\mathbf{p})$ is the drag coefficient representing the thermal relaxation rate ($T$: medium temperature).
The momentum diffusion coefficient $D(\mathbf{p})$ is obtained from $A(\mathbf{p})$ via the Einstein relation.

The pertinent numerical realization of the Fokker--Planck equation is then given by a set of stochastic Langevin equations~\cite{He:2013zua} as:
\begin{equation}
\begin{aligned}
\dd x_i &= \frac{p_i}{E}\,\dd t\,,\\
\dd p_i &= -\Gamma(\mathbf{p})\,p_i\,\dd t
+ \sqrt{2\,D\!\left(|\mathbf{p}+\dd\mathbf{p}|\right)\,\dd t}\;\rho_i \, ,
\end{aligned}
\end{equation}
where $E=\sqrt{\mathbf{p}^2+m_c^2}$ denotes the charm quark's energy and
$x_i$ and $p_i$ ($i=1,2,3$) are its spatial and momentum coordinates, respectively.
The stochastic variables $\rho_i$ are Gaussian-distributed random numbers
satisfying
$\langle \rho_i \rangle = 0$ and $\langle \rho_i \rho_j \rangle = \delta_{ij}$,
with a probability distribution
$P(\rho) = (2\pi)^{-3/2}\exp(-\boldsymbol{\rho}^2/2)$.
The coordinate update is straightforward using the velocity determined by the current momentum.
The momentum update is carried out in the local rest frame of the medium,
in which the drag and diffusion coefficients are defined.
First, a provisional momentum increment is computed using the drag and
diffusion coefficients evaluated at the initial momentum $\mathbf{p}$.
Subsequently, the diffusion term is evaluated at the updated momentum
magnitude $|\mathbf{p}+\dd\mathbf{p}|$,
\begin{equation}
\dd p_i^{\mathrm{diff}}
=
\sqrt{2\,D\!\left(|\mathbf{p}+\dd\mathbf{p}|\right)\,\dd t}\;\rho_i \, ,
\end{equation}
and combined with the drag contribution,
$\dd p_i^{\mathrm{drag}}=-\Gamma(\mathbf{p})\,p_i\,\dd t$,
to obtain the total momentum increment $\dd p_i$.
The same time step $\dd t$ and stochastic noise $\rho_i$ are used in both steps. After each time step, the updated momentum is Lorentz-boosted back to the laboratory frame.

\subsection{Relaxation rate}
\label{ssec:A}
The key input to the HQ Langevin simulation is the relaxation rate, which
encodes the interaction strength between the heavy quark and the medium
constituents. We calculate it using nonperturbative $T$-matrices with recent constraints from WLCs computed in lattice QCD~\cite{Tang:2023tkm}. Of particular importance are off-shell effects that enable the $c$-quark to access interaction strength from $D$-meson bound states that are located below the nominal charm light-quark threshold.
The pertinent expression has been derived from the Kadanoff-Baym equations in Ref.~\cite{Liu:2018syc}. For an incoming on-shell $c$-quark of momentum $\mathbf{p}_c$ one obtains
\begin{equation}
   \begin{aligned}
& A(\mathbf{p}_c) =
\sum_{i} 
\int \dd \mathbf{\Pi}\,
\frac{1}{d_c}
|\mathcal{M}_{ic\rightarrow ic}|^2
\\
& \quad \times \rho_c(\omega', \mathbf{p}_c') \,
\rho_i(\nu, \mathbf{q}) \,
\rho_i(\nu', \mathbf{q}')
\\
& \quad \times [1 - n_c(\omega')] \,
n_i(\nu) \,
[1 \pm n_i(\nu')]
\left(1 - \frac{\mathbf{p}_c \cdot \mathbf{p}_c'}{\mathbf{p}_c^2}\right)\,,
\label{eq:A_p}
   \end{aligned}
\end{equation}
where the off-shell effects are encoded in the heavy-light scattering amplitude, 
$\mathcal{M}_{ic\rightarrow ic}$, and in the 
spectral functions, $\rho_{i,c}$,  of the light thermal partons ($i = u, \bar{u}, d, \bar{d}, s, \bar{s}, g$ with 4-momenta $(\nu, \mathbf{q})$ and $(\nu', \mathbf{q}')$ for the incoming and outgoing ones, respectively) and the outgoing $c$-quark of momentum $\mathbf{p}_c'$
(for the incoming $c$-quark, on-shell kinematics are sufficient within the semiclassical Langevin framework employed here).
The corresponding on-shell energies, $\varepsilon_i(\mathbf{q})=\sqrt{m_i^{2}+\mathbf{q}^2}$
and $\varepsilon_c(\mathbf{p}_c)=\sqrt{m_c^{\,2}+\mathbf{p}_c^2}$, figure in the integration measure,
\begin{equation}
    \begin{aligned}
\dd \mathbf{\Pi} &=  
\frac{\dd\omega' \, \dd^3 \mathbf{p}_c'}{(2\pi)^3 2 \varepsilon_c(\mathbf{p}_c')}
\frac{\dd\nu \, \dd^3 \mathbf{q}}{(2\pi)^3 2 \varepsilon_i(\mathbf{q})}
\frac{\dd\nu' \, \dd^3 \mathbf{q}'}{(2\pi)^3 2 \varepsilon_i(\mathbf{q}')}
\\
& \times (2\pi)^4
\delta^{(3)} (\mathbf{p}_c + \mathbf{q} - \mathbf{p}_c' - \mathbf{q}')
\\
& \times \frac{1}{2 \varepsilon_c(\mathbf{p}_c)} \delta ( \varepsilon_c(\mathbf{p}_c) + \nu - \omega' - \nu')\,, 
    \end{aligned}
    \label{eq:measure}
\end{equation}
which also contains the energy-momentum conserving $\delta$-functions. 
The $c$-quark's color--spin degeneracy is $d_c$=6, and $n$ is the Fermi--Dirac  
(Bose--Einstein) distribution for fermions (bosons).   

The heavy-light scattering amplitude,
$|\mathcal{M}_{ic\rightarrow ic}|^2$, is directly related to the $T$-matrix which we evaluate in the center-of-mass frame, including the sum over all relevant
color and partial-wave channels~\cite{Liu:2018syc}.
For later use we introduce the spin-color averaged matrix element,
\begin{equation}
\overline{|\mathcal{M}_{ic\rightarrow ic}|^2}
= \frac{|\mathcal{M}_{ic\rightarrow ic}|^2}{d_i d_c}\,,
\end{equation}
where $d_i$ denotes the color-spin degeneracy of the light partons, 
$d_{q/\bar q}=6$ and $d_g=16$.

\begin{figure}[t]
    \centering
    \includegraphics[width=0.47\textwidth]{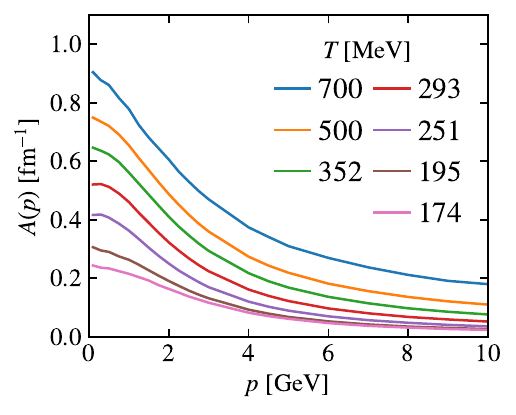}
     \caption{The charm-quark relaxation rate, $A(p)$, as a function of momentum at different temperatures from $T$-matrices constrained by WLCs from lattice QCD~\cite{Tang:2023tkm}.}
    \label{fig:A_of_p}
\end{figure}
In Fig.~\ref{fig:A_of_p} we show the charm-quark relaxation rate as a function of 3-momentum, $p=|\mathbf{p}_c|$, at different temperatures.
Its momentum dependence features a marked enhancement at low momenta where nonperturbative effects are most significant, which was previously mimicked by perturbative reaction rates with a large $K$ factor~\cite{Du:2022uvj}.

In our comparisons to charmonium data in Sec.~\ref{sec:lhc} we will also encounter a so-called non-prompt contribution from final-state decays of bottom hadrons (referred to as $B$ feeddown), whose $\pT$ spectrum is affected by the rescattering (and hadronization) of bottom quarks in the fireball. We take this effect into account by also conducting Langevin simulations of $b$-quarks in the QGP with transport coefficients that are derived from the same $T$-matrix interactions as for the $c$-quarks, using the same methods (\eg, off-shell effects).
For simplicity we will then estimate the momentum-dependent $B$-feeddown contribution by multiplying the non-prompt
charmonium spectrum from $pp$ collisions~\cite{LHCb:2021pyk} by the $b$-quark $\raa$ (neglecting effects of $b$-quark hadronization and detailed decay kinematics).

\section{Charmonium transport}
\label{sec:boltzmann}
In this section, we outline the transport framework for charmonia in the QGP. In Sec.~\ref{ssec:boltzmann}, we introduce the Boltzmann transport equation and its formal solution, which separates into primordial and regeneration components. In Sec.~\ref{ssec:rates}, we construct the nonperturbative dissociation (Sec.~\ref{sssec_alpha}) and regeneration rates (Sec.~\ref{sssec_beta}) and elucidate their connection to charm-quark transport, thereby verifying 
the proper detailed balance relation in the limit of thermalized charm quarks. 
This requires a discussion of the charmonium equilibrium limit in the presence of off-shell effects, which is given in Sec.~\ref{ssec:equilibrium_limit}. 

\subsection{Boltzmann equation}
\label{ssec:boltzmann}
The transition between bound charmonium states, $\Psi=J/\psi, \chi_c, \psi(2S)$, 
and uncorrelated charm-anticharm pairs in the medium can be evaluated in semiclassical approximation using the Boltzmann transport equation,
\begin{equation}
\frac{\partial f_{\Psi}}{\partial t}
+ \mathbf{v}\cdot\nabla f_{\Psi}
= -\alpha(\mathbf{P}_{\Psi},T)\, f_{\Psi}
+ \beta(\mathbf{P}_{\Psi},T)\,,
\label{eq:boltzmann}
\end{equation}
where
\begin{equation}
f_{\Psi}(\mathbf{x},\mathbf{P}_{\Psi},t)
\equiv \frac{\mathrm{d}N_{\Psi}}{\mathrm{d}^3x\,\mathrm{d}^3P_{\Psi}}\,,
\end{equation}
is the phase-space distribution function of a charmonium state of
momentum $\mathbf{P}_{\Psi}$ at position $\mathbf{x}$, and its velocity $\mathbf{v}=\mathbf{P}_{\Psi}/E_{\Psi}$ with energy $E_{\Psi}=\sqrt{\mathbf{P}_{\Psi}^2+M_{\Psi}^2}$. 

The formal solution of Eq.~\eqref{eq:boltzmann} amounts to a sum of a
primordial-survival contribution and a continuous regeneration contribution~\cite{PhysRevLett.97.232301},
\begin{equation}
\begin{aligned}
&f(\mathbf{x},\mathbf{P}_{\Psi},t)
=
f\!\left(\mathbf{x}(t,t_0),\mathbf{P}_{\Psi},t_0\right)
e^{-\int_{t_0}^{t}\! \mathrm{d}t'\,
\alpha(\mathbf{P}_{\Psi},T(t'))}
\\
&\qquad+
\int_{t_0}^{t}\! \mathrm{d} t'\,
\beta(\mathbf{P}_{\Psi},T(t'))
e^{-\int_{t'}^{t}\! \mathrm{d}t''\,
\alpha(\mathbf{P}_{\Psi},T(t''))}\,,
\end{aligned}
\label{eq:boltzmann_solution}
\end{equation}
where $\mathbf{x}(t,t') \equiv \mathbf{x}-\mathbf{v}(t-t')$ represents the classical
charmonium trajectory.
The first term in Eq.~\eqref{eq:boltzmann_solution} tracks the primordial charmonium component produced from initial hard scatterings. When propagating through the evolving medium, it undergoes a gradual dissociation governed by the in-medium rate $\alpha$ (we neglect elastic rescattering, which is suppressed by the bound-state phase space).
The second term accounts for charmonium regeneration from 
charm-anticharm pairs, which commences at the dissociation temperature and continues until freezeout. The incremental gain, $\beta\,\mathrm{d}t$, is produced locally in the medium and thus inherits the collective flow of the surrounding fluid (imparted on the charm quarks), and subsequently it is subject to the same dissociation processes as the primordial component.
The time integral therefore reflects the competition between continuous regeneration and subsequent in-medium suppression, driving the total yields towards their equilibrium.



\subsection{Nonperturbative reaction rates}
\label{ssec:rates}
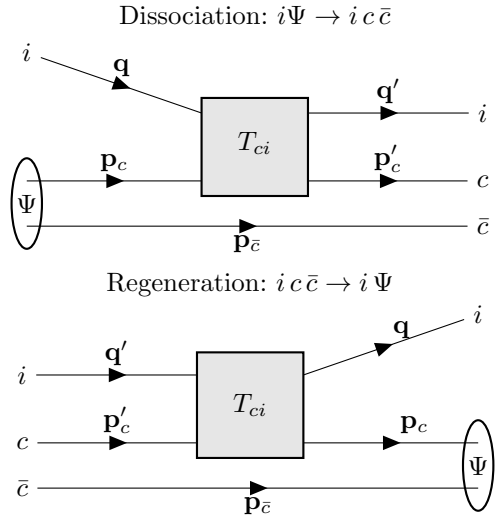
\begin{figure}[t]
\centering
\begin{minipage}{0.48\textwidth}
\centering
\begin{tikzpicture}
   \begin{feynman}
      \vertex (i) at (-3,2) {\(i\)};
      \vertex (Qi) at (-3,0.3);
      \vertex (Qbari) at (-3,-0.3);
      \vertex (CalQ) at (-3,0) {\(\Psi\)};
      \coordinate (rectL1) at (-0.7,1.2);
      \coordinate (rectL2) at (-0.7,0.3);
      \coordinate (rectR1) at (0.7,1.2);
      \coordinate (rectR2) at (0.7,0.3);
      \vertex (iprime) at (3,1.2) {\(i\)};
      \vertex (Q) at (3,0.3) {\(c\)};
      \vertex (Qbar) at (3,-0.3) {\(\bar{c}\)};

      \draw [thick] (CalQ) ellipse [x radius=0.2cm, y radius=0.6cm];
      \draw [thick, fill=gray!20] (-0.7,0.1) rectangle (0.7,1.4);
      \node at (0,0.8) {\({T}_{ci}\)};

      \draw[fermion] (i) -- (rectL1) node[midway, above] {\({\mathbf{q}}\)};
      \draw[fermion] (Qi) -- (rectL2) node[midway, above] {\({\mathbf{p}}_c\)};
      \draw[fermion] (rectR1) -- (iprime) node[midway, above] {\({\mathbf{q}}'\)};
      \draw[fermion] (rectR2) -- (Q) node[midway, above] {\({\mathbf{p}}_c^{\prime}\)};
      \draw[fermion] (Qbari) -- (Qbar) node[midway, below] {\({\mathbf{p}}_{\bar{c}}\)};
   \end{feynman}
   \node at (0,2.5) {Dissociation: \( i \Psi \rightarrow i\, c\, \bar{c} \)};
\end{tikzpicture}
\end{minipage}
\hfill
\begin{minipage}{0.48\textwidth}
\centering
\begin{tikzpicture}
   \begin{feynman}
      \vertex (i) at (-3,1.2) {\(i\)};
      \vertex (Q) at (-3,0.3) {\(c\)};
      \vertex (Qbar) at (-3,-0.3) {\(\bar{c}\)};
      \coordinate (rectL1) at (-0.7,1.2);
      \coordinate (rectL2) at (-0.7,0.3);
      \coordinate (rectR1) at (0.7,1.2);
      \coordinate (rectR2) at (0.7,0.3);
      \vertex (iprime) at (3,2.0) {\(i\)};
      \vertex (CalQ) at (3,0) {\(\Psi\)};
      \vertex (CalQtop) at (3,0.3);
      \vertex (CalQbot) at (3,-0.3);

      \draw [thick] (CalQ) ellipse [x radius=0.2cm, y radius=0.6cm];
      \draw [thick, fill=gray!20] (-0.7,0.1) rectangle (0.7,1.5);
      \node at (0,0.8) {\({T}_{ci}\)};

      \draw[fermion] (i) -- (rectL1) node[midway, above] {\({\mathbf{q}}'\)};
      \draw[fermion] (Q) -- (rectL2) node[midway, above] {\({\mathbf{p}}_c^{\prime}\)};
      \draw[fermion] (rectR1) -- (iprime) node[midway, above right] {\({\mathbf{q}}\)};
      \draw[fermion] (rectR2) -- (CalQtop) node[midway, above right] {\({\mathbf{p}}_c\)};
      \draw[fermion] (Qbar) -- (CalQbot) node[midway, below] {\({\mathbf{p}}_{\bar{c}}\)};
   \end{feynman}
   \node at (0,2.4) {Regeneration: \( i\, c\, \bar{c} \rightarrow i\,\Psi \)};
\end{tikzpicture}
\end{minipage}
\caption{
Schematic $T$-matrix diagrams illustrating the dissociation and regeneration of charmonium in the QGP.
\textit{Upper panel}: dissociation process \(i\,\Psi \to i\,c\bar c\), in which a thermal parton \(i\) scatters off a bound charmonium state \(\Psi\), leading to its breakup into an unbound charm--anticharm pair.
\textit{Lower panel}: inverse process of regeneration \(i\,c\bar c \to i\,\Psi\), where an uncorrelated \(c\bar c\) pair recombines into charmonium through an interaction with a thermal parton.
In both cases, the interaction is described by the half-off-shell \(T\)-matrix amplitude \(T_{ci}\).
The 3-momenta of the thermal partons and heavy quarks are indicated explicitly.
}
\label{fig:diss-reg}
\end{figure}

In Fig.~\ref{fig:diss-reg} we pictorially illustrate the dissociation and regeneration mechanisms of charmonia in the QGP. We employ a quasifree approximation~\cite{Grandchamp:2003uw,Wu:2025hlf}, where the
$2\!\to\!3$ dissociation process (upper panel) 
is reduced to an inelastic (half off-shell) $2\!\to\!2$ scattering process
of a thermal-medium parton off the charm (anticharm) quark, while the anticharm (charm) quark acts as a spectator.
The charmonium binding energy is incorporated
into an effective charm-quark mass,
\begin{equation}
\tilde m_c = m_c - E_B^{\Psi} \ .
\end{equation}
The in-medium charm-quark mass, $m_c$, and charmonium binding energies, $E_B^\Psi$, are determined from the same in-medium $T$-matrix formalism as the heavy-light amplitudes~\cite{Tang:2023tkm}.
For a charmonium 3-momentum $\mathbf{P}_{\Psi}$ the momenta
of the charm and anticharm quarks are assigned proportionally to their
effective masses,
\begin{equation}
\label{eq:qfmass}
\mathbf{p}_c = \frac{\tilde m_c}{M_{\Psi}} \mathbf{P}_{\Psi}\,,
\qquad
\mathbf{p}_{\bar c} = \frac{m_c}{M_{\Psi}} \mathbf{P}_{\Psi}\,,
\end{equation}
which ensures equal velocities and preserves relativistic energy-momentum conservation.
The charmonium energy is simply the sum of the energies of the quasifree charm quark and the spectator anticharm quark,
\begin{equation}
\label{eq:qf_E}
E_{\Psi}(\mathbf{P}_{\Psi}) = \varepsilon_c(\mathbf{p}_c)
+ \varepsilon_{\bar c}(\mathbf{p}_{\bar c})\,.
\end{equation}
Energy conservation in the
$2\!\to\!3$ ($3\!\to\!2$) process reads
\begin{equation}
\label{eq:energy_conservation}
E_{\Psi}(\mathbf{P}_{\Psi}) + \nu
=
\omega' + \nu' + \varepsilon_{\bar c}(\mathbf{p}_{\bar c}) \, ,
\end{equation}
and reduces to
\begin{equation}
\varepsilon_c(\mathbf{p}_c) + \nu
=
\omega' + \nu' \, ,
\end{equation}
for the $2\!\to\!2$ process. This equivalence is critical for recovering the correct equilibrium limit, as will be elaborated in the derivation of the regeneration rate later in this section.

\begin{figure}[tbp]
    \centering
   \hspace{-0.35cm} 
   \includegraphics[width=0.49\textwidth]{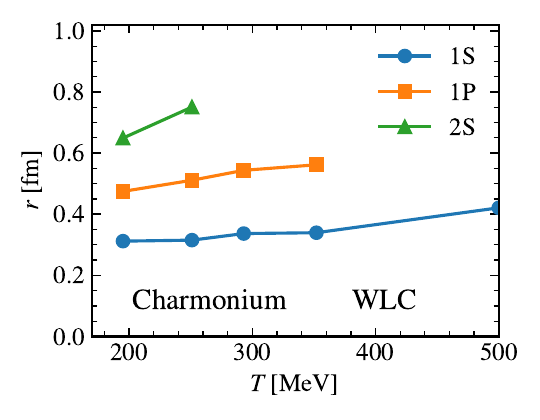}
    \caption{The in-medium radii of charmonia as a function of temperature, obtained from matching the calculated dissociation rates to the (imaginary part of the) pole positions from in-medium quarkonium $T$-matrices~\cite{Tang:2025ypa} as described in Ref.~\cite{Wu:2025hlf}. 
    }
    \label{fig:radius}
\end{figure}

\begin{figure*}[htbp]
    \centering
    \includegraphics[width=0.99\textwidth]{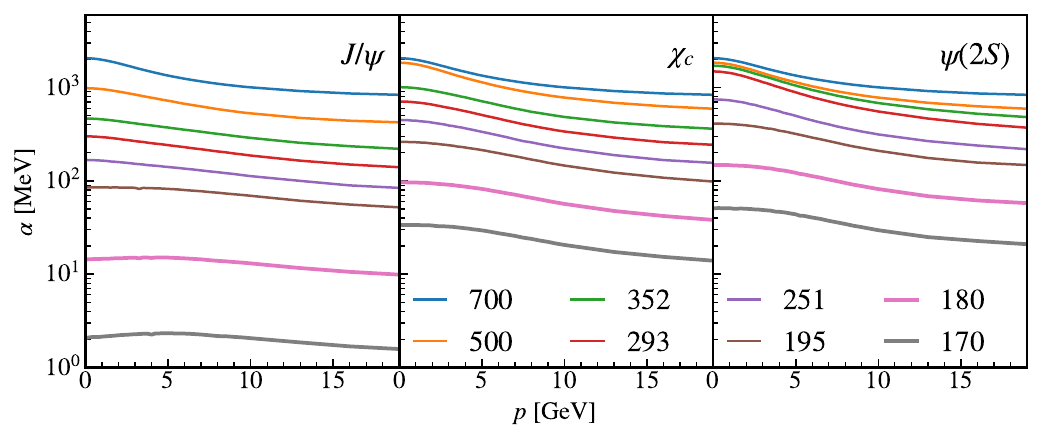}

    \caption{Dissociation rates of three charmonium states calculated within the nonperturbative $T$-matrix formalism in the WLC scenarios, shown as functions of their momenta at different temperatures (different colors).
    From left to right, the panels correspond to $J/\psi$, $\chi_c$, and $\psi(2S)$.
    }
    \label{fig:alpha}
\end{figure*}

\subsubsection{Dissociation rate}
\label{sssec_alpha}
The final expression for the charmonium dissociation rate is given by~\cite{Liu:2018syc,Wu:2025hlf}
\begin{equation}
\label{eq:alpha}
\begin{aligned}
&\alpha(\mathbf{P}_{\Psi}, T(t)) = 2 \sum_i d_i
\int \dd \mathbf{\Pi}\;
\overline{|\mathcal{M}_{ic\to ic}|^2} \\
&\qquad\times
\rho_c(\omega', \mathbf{p}_c')\,
\rho_i(\nu, \mathbf{q})\,
\rho_i(\nu', \mathbf{q}') \\
&\qquad\times
\left(1 - n_c(\omega')\right)
n_i(\nu)
\left(1 \pm n_i(\nu')\right)
\left(1 - n_{\bar c}(\varepsilon_{\bar c})\right) \\
&\qquad\times
\left[1 - e^{i \mathbf{k}\cdot \mathbf{r}}\right] \, ,
\end{aligned}
\end{equation}
where the sum runs over thermal partons $i=q,\bar q,g$.
The phase-space integration measure $\dd\mathbf{\Pi}$ is defined in Eq.~\eqref{eq:measure}, but with a "quasifree" (off-shell) incoming charm quark.
%
The scattering matrix and spectral functions are identical to those in Eq.~\eqref{eq:A_p}, but with the addition of a Pauli-blocking factor for the
spectator anticharm quark (which is numerically immaterial).
The factor $1 - e^{i\mathbf{k}\cdot\mathbf{r}}$, motivated by its perturbative expression from Ref.~\cite{Laine:2006ns}, accounts for the interference
between charm--light and anticharm--light scattering amplitudes,
where $\mathbf{k}=\mathbf{p}_c'-\mathbf{p}_c$ denotes the momentum transfer and
$\mathbf{r}$ characterizes the spatial size of the charmonium bound state, see Fig.~\ref{fig:radius}~\cite{Wu:2025hlf}.
We recall that the temperature-dependent radii have been adjusted so that the dissociation rate at vanishing charmonium momentum agrees with the width from the full $T$-matrix calculations (\ie, from their pole positions in the complex plane). In this way the full structure information, including the relative motion of the anti-/charm quarks inside the bound states, is effectively accounted for in the above rate expression.
For large charmonium radii, the rapidly oscillating phase suppresses the interference term, effectively turning off interference effects, while destructive interference causes the rate to vanish in the limit $r\to0$.

In Fig.~\ref{fig:alpha} we collect the resulting rates as a function of 3-momentum and for different temperatures for the three charmonium states considered in this work. A clear hierarchy is observed in terms of the binding energy of the
different states with smaller values leading to larger dissociation widths (rates).
Compared to the perturbative dissociation rates employed in our previous
works~\cite{Wu:2024gil}, the nonperturbative rates are substantially larger,
especially for the excited states. Moreover, the momentum dependence of the nonperturbative rates differs
qualitatively from that of the perturbative calculations: while the
nonperturbative rates decrease with increasing momentum, the leading-order
perturbative rates exhibit an increasing trend with momentum.
We also make use of a $T$-matrix pole analysis of the in-medium charmonium spectrum in the complex energy plane~\cite{Tang:2025ypa}, which indicates that the $J/\psi$, $\chi_c$, and $\psi(2S)$ states persist up to temperatures of about 500, 350 and 250\,MeV, respectively. These temperatures, which are generally higher than where the binding energies vanish, are used to determine the onset of the regeneration processes in the transport simulations.

\begin{figure}[htbp]
    \centering
    \hspace{-0.35cm} 
    \includegraphics[width=0.49\textwidth]{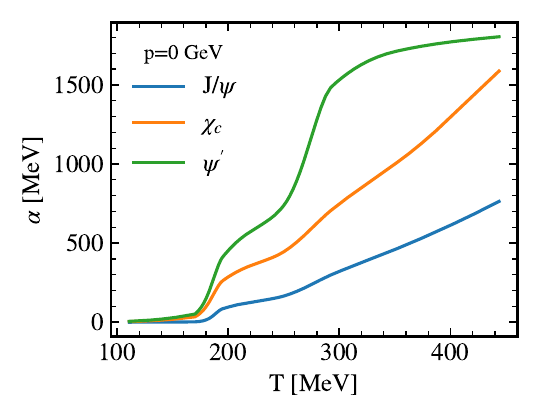}
    
    \caption{Charmonium dissociation rates at vanishing three-momentum
    as a function of temperature.
    }
    \label{fig:alpha_p0}
\end{figure}

The reaction rates for charmonia in confined matter are generally believed to be rather small due to small bound-state sizes and the absence of a constituent light-quark content. For the $J/\psi$ this is well supported by a wide variety of effective-model calculations, see, \eg, Ref.~\cite{Rapp:2008tf}. We here employ hadronic calculations of Ref.~\cite{Du:2015wha} where $SU(4)$-flavor based rates for $\pi$- and $\rho$-induced reactions~\cite{Lin:1999ad,Haglin:2000ar} have been extended to a hadron resonance gas using phase space scaling. We then interpolate our QGP-based rates at $T$=195~MeV down to these hadronic rates at 170~MeV at $p=0$~GeV~\cite{Du:2015wha}.
The momentum dependence at 170~MeV is inherited from the corresponding
$T$-matrix calculation at 174~MeV.
For the excited states, the hadronic rates have a hierarchy similar to the lowest QGP temperature, but they are more suppressed for the $J/\psi$. On the other hand, the open-charm thermalization rates (plotted in Fig.~\ref{fig:A_of_p}) are much less suppressed toward the hadronic phase, which is compatible with the phenomenological observation of a sequential chemical and thermal freezeout in heavy-ion collisions (\ie, chemical reactions shut off well before elastic interactions).
In Fig.~\ref{fig:alpha_p0} we summarize the temperature dependence of the rates at vanishing charmonium momentum. They nicely exhibit the expected hierarchy based on their vacuum binding properties which not only includes the binding energy but also interference effects due to their different sizes which increasingly suppress the rates at smaller radii (also referred to as the imaginary part of the potential). 
When the state dissolves, the rates level off at twice the collision rate of the constituent charm quarks; toward lower temperatures the interference effects kick in first followed by another drop induced by the transition into the hadronic phase, while the reduction due to finite binding energies is much more gradual (and strongly smeared out due to the large widths of the charm-quark spectral functions).

\subsubsection{Regeneration rate}
\label{sssec_beta}
We now turn to the regeneration rate, $\beta$, which, to our knowledge, is the first time that it has been worked out with off-shell spectral functions.
We start from the detailed-balance relation in thermal equilibrium~\cite{Du:2022uvj},
\begin{equation}
\label{eq:detail_balance}
\beta(\mathbf{P}_{\Psi},T)
=
\gamma_c^2\, d_{\Psi}\,
e^{-E_{\Psi}(\mathbf{P}_{\Psi})/T}\,
\alpha(\mathbf{P}_{\Psi},T)\, ,
\end{equation}
where $\gamma_c$ is the charm-quark fugacity that accounts for charm-number conservation in the medium~\cite{Zhao:2010nk,Du:2017qkv}, and $d_{\Psi}$ is the spin degeneracy of each charmonium state (we use 8 for the near-degenerate $\chi_c(1P)$ states).
By energy conservation in Eq.~\eqref{eq:energy_conservation}, the Boltzmann factor takes the form
\begin{equation}
\label{eq:identity1}
e^{-E_{\Psi}(\mathbf{P}_{\Psi})/T}
=
e^{-(\omega' + \nu' - \nu + \varepsilon_{\bar c})/T} \, .
\end{equation}
We further employ the identity,
\begin{equation}
\label{eq:identity2}
n(E) = e^{-E/T}\bigl(1 \pm n(E)\bigr) \, ,
\end{equation}
(where the upper (lower) sign applies to bosons (fermions)) which allows us
to rewrite Boltzmann factors into quantum-statistical distribution functions
and their pertinent Bose-enhancement or Pauli-blocking factors.
Combining Eqs.~\eqref{eq:alpha}, \eqref{eq:detail_balance},
\eqref{eq:identity1}, and \eqref{eq:identity2}, we then obtain the regeneration rate in thermal equilibrium as
\begin{equation}
\label{eq:beta_thermal}
\begin{aligned}
&\beta(\mathbf{P}_{\Psi},T)
=
2\,\gamma_c^2
\sum_i d_{\Psi} d_i
\int \dd \mathbf{\Pi}\;
\overline{|\mathcal{M}_{ic\to ic}|^2} 
\\
&\qquad\qquad \times
\rho_c(\omega',\mathbf{p}_c')\,
\rho_i(\nu,\mathbf{q})\,
\rho_i(\nu',\mathbf{q}')
\\
&\qquad\qquad \times
n_c(\omega')\,
[1\pm n_i(\nu)]\,
n_i(\nu')\,
n_{\bar c}(\varepsilon_{\bar c})
\\
&\qquad \qquad \times
\left[1 - e^{i\mathbf{k}\cdot\mathbf{r}}\right]\, 
\end{aligned}
\end{equation}
where all distribution functions correspond to thermal equilibrium.
For dissociation, the incoming $c$-quark is taken to be on shell, while
the incoming and outgoing light partons $i$ and the outgoing $c$-quark
are treated off shell.
For regeneration, where the quarkonium appears in the final state, the
on-shell and off-shell assignments are interchanged relative to dissociation:
the outgoing $c$-quark is taken as on-shell, whereas the incoming heavy quark and
the incoming and outgoing light partons $i$ remain off-shell.
The spectator quark is on-shell in both rates.

Off-equilibrium charm quarks can now be implemented into the regeneration processes
by replacing the charm-quark distribution functions, $n_c$, with $f_c(p,t)$ from the Langevin simulations. This is straightforward for the on-shell distribution function,
$n_{\bar c}(\varepsilon_{\bar c})$, where the on-shell energy is directly available from the semiclassical Langevin simulation at each time step. But it requires some care for the off-shell case, $n_c(\omega')$. It turns out that an additional factor  
$e^{-[\omega' - \varepsilon_c(\mathbf{p}_c')]/T(t)}$, which essentially represents
a thermal off-shell weighting, 
results in the correct equilibrium limit.
The thermal off-equilibrium and off-shell regeneration rate is thus given by
\begin{equation}
\label{eq:beta}
\begin{aligned}
&\beta(\mathbf{P}_{\Psi},T(t))
=
2\,\gamma_c^2
\sum_i d_{\Psi} d_i
\int \dd \mathbf{\Pi}\;
\overline{|\mathcal{M}_{ic\to ic}|^2}
\\
&\qquad \qquad \qquad \times
\rho_c(\omega',\mathbf{p}_c')\,
\rho_i(\nu,\mathbf{q})\,
\rho_i(\nu',\mathbf{q}')
\\
&\qquad \qquad \qquad \times
[1\pm n_i(\nu)]\,
n_i(\nu')
\\
&\qquad \qquad \qquad \times
f_c(\mathbf{p}_c',T(t))\,
e^{-[\omega' - \varepsilon_c(\mathbf{p}_c')]/T(t)}
\\
&\qquad \qquad \qquad \times
f_{\bar c}(\mathbf{p}_{\bar c},T(t))\,
\left[1 - e^{i\mathbf{k}\cdot\mathbf{r}}\right]\,.
\end{aligned}
\end{equation}
In the limit where charm quarks are fully thermalized, the non-equilibrium
distribution becomes
\begin{equation}
f_c(\mathbf{p}_c',T(t))
\;\to\;
e^{-\varepsilon_c(\mathbf{p}_c')/T(t)} \, ,
\end{equation}
which renders the Boltzmann factor in Eq.~\eqref{eq:beta}
\begin{equation}
f_c(\mathbf{p}_c',T(t))\,
e^{-[\omega' - \varepsilon_c(\mathbf{p}_c')]/T(t)}
\;\longrightarrow\;
e^{-\omega'/T(t)} \,,
\end{equation}
thus recovering Eq.~\eqref{eq:beta_thermal}.


\subsection{Equilibrium limit}
\label{ssec:equilibrium_limit}
The equilibrium limit of charmonia plays a key role in their kinetics. Since charm-quark number is conserved, it depends on the interplay with the open-charm spectrum in the medium which is encoded in the fugacity factor (or chemical potential) introduced in Eq.~\eqref{eq:detail_balance}. The relative chemical equilibrium in the charm-quark sector thus depends on the charm-quark masses and charmonium binding energies which in turn determine the charmonium masses.
This can be directly gleaned from the stationary (long-time) limit of the Boltzmann equation, Eq.~\eqref{eq:boltzmann},
\begin{equation}
\frac{\mathrm{d} f_{\Psi}}{\mathrm{d} t} = 0 \,,
\end{equation}
which leads to the detailed-balance condition
\begin{equation}\label{eq:eq_limit}
    f_{\Psi}^{\rm eq}(\mathbf{x},\mathbf{p},t)
    = \frac{\beta(\mathbf{x},\mathbf{p},t)}{\alpha(\mathbf{x},\mathbf{p},t)} \,,
\end{equation}
and thus 
\begin{equation}
    f_{\Psi}^{\rm eq}(\mathbf{x},\mathbf{p},t)
    = \gamma_c^2\, d_{\Psi}\, e^{-E_{\Psi}/T(t)} \,.
\end{equation}

\begin{figure}[t]
    \centering
    \includegraphics[width=0.47\textwidth]{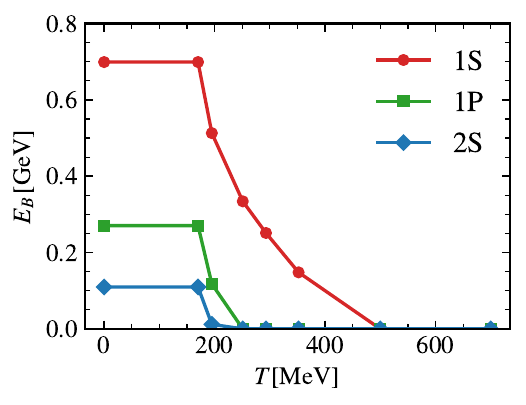}
     \caption{Binding energies of $J/\psi$, $\chi_c$ and $\psi(2S)$ as a function of temperature, extracted from the in-medium $T$-matrix approach in the WLC scenario.}
    \label{fig:eb}
\end{figure}
\begin{figure}[t]
    \centering
    \includegraphics[width=0.47\textwidth]{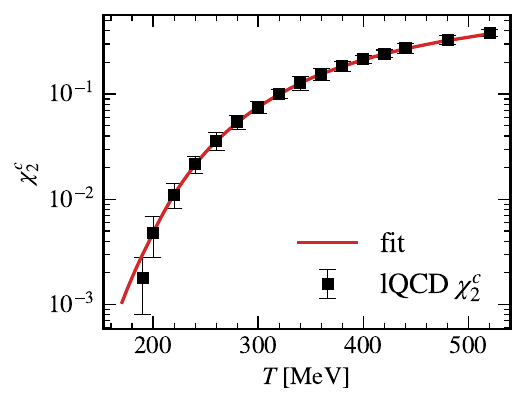}
    \includegraphics[width=0.47\textwidth]{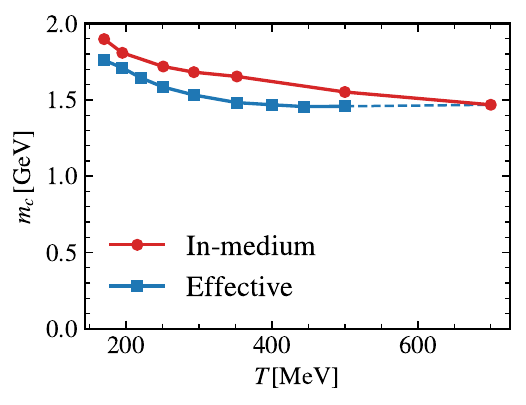}
    \caption{
        Upper panel: Charm-quark number susceptibility from lattice-QCD calculations~\cite{Bellwied:2015lba} compared to the quasiparticle model with an effective charm-quark mass.
        Lower panel: In-medium charm-quark mass extracted from the $T$-matrix in the WLC scenario, together with the effective charm-quark mass, as functions of temperature.
        }
    \label{fig:mc_eff_T}
\end{figure}

The in-medium charm-quark mass and charmonium binding energies are taken from the thermodynamic $T$-matrix constrained by the lattice-QCD results of the WLCs~\cite{Wu:2025hlf} and are shown in Figs.~\ref{fig:eb} and \ref{fig:mc_eff_T}, respectively.

The charm-quark fugacity, $\gamma_c$, is fixed by the total number of produced 
charm-quark pairs, $N_{c\bar{c}}$, in the fireball of a heavy-ion collision, 
\begin{equation}
    N_{c\bar{c}} = \frac{1}{2} \gamma_c n_{\rm op} V_{\rm FB} \frac{I_1(\gamma_c n_{\rm op} V_{\rm co})}{I_0(\gamma_c n_{\rm op} V_{\rm co})} + \gamma_c^2 n_{\rm hid} V_{\rm FB}\,,
    \label{eq:fugacity}
\end{equation}
where $V_{\rm FB}$ is the (time-dependent) fireball volume,
and $n_{\rm op}$ and $n_{\rm hid}$ denote the densities of open and hidden charm states at a given temperature, respectively (the latter is usually negligible); $I_0$ and $I_1$ are the modified Bessel functions;
$V_{\rm co}$ denotes the correlation volume that accounts for the local conservation of charm number in the canonical ensemble~\cite{Hamieh:2000tk,Grandchamp:2003uw,Zhao:2010nk}, which is particularly relevant if only one (or a few) pair(s) are present, which effectively increases the local $c\bar{c}$ density. We follow our previous treatment which accounts for a time-dependent expansion as 
$V_{\rm co}(\tau) = \frac{4\pi}{3} \left(r_0 + \langle v_c \rangle \tau\right)^3$
where $r_0 \simeq 1.2\,$fm is an initial radius and $\langle v_c \rangle = 0.6$ is a typical recoil velocity of the charm quarks. 
When the system contains multiple charm-quark pairs, their correlation volumes overlap 
and the system transitions to the grand-canonical ensemble, \ie, the correlation volume is replaced by the fireball volume, $V_{\rm co} \rightarrow V_{\rm FB}$. 
In this work, we use the condition $N_{c\bar{c}} V_{\rm co} > \kappa V_{\rm FB}$ to switch from the canonical to the grand-canonical ensemble.
Our default value $\kappa$=1 renders a continuous transition, but we have checked that the regeneration yields in Pb--Pb collisions are hardly affected by varying $\kappa$ in either semi-/central collisions (where the abundant charm production puts the system into the grand-canonical regime already at early times) or peripheral collisions (where the average $c\bar{c}$ number is well below 1 so that the system remains in the canonical regime throughout the evolution). 
The main sensitivity arises at intermediate centralities (around $N_{\rm part}\simeq 100$), where typically 1--2 $c\bar{c}$ pairs are present: \eg, reducing $\kappa$ from 1 to 0.5 leads to an earlier transition to the grand-canonical ensemble, lowering the equilibrium limit and reducing the regenerated charmonium yield by about 10--20\%.

The open-charm density, $n_{\rm op}$, in the QGP phase was previously estimated by including both the charm quarks and the lowest-lying $S$-wave $D$-meson states ($D$, $D^{*}$, $D_{s}$, and $D_{s}^{*}$)~\cite{Zhao:2010nk}. In this work, we instead employ only charm quarks but with an effective mass, $m_c^*(T)$, that is constrained by fitting the lattice-QCD results for the (dimensionless) charm-quark number susceptibility~\cite{Bellwied:2015lba,Kaczmarek:2025dqt}:
\begin{equation}
\chi_2^{\mathrm{c}}
=
\frac{2 d_c}{(2\pi T)^3}
\int \dd^3\mathbf{p}\, e^{-E_p/T}\, ,
\end{equation}
with $E_p = \sqrt{p^2 + m_c^{*2}(T)}$, $d_c$=6 and an extra factor 2 to account for charm and anticharm quarks. By fitting the lattice-QCD data for the charm-quark number susceptibility, shown in the upper panel of Fig.~\ref{fig:mc_eff_T}, we extract the temperature dependence of the effective charm-quark mass, which is displayed in the lower panel of Fig.~\ref{fig:mc_eff_T}.
At a temperature of $T_{\rm H} = 170\,$MeV, where we convert from partonic to hadronic degrees of freedom in our transport simulations, we have ensured that the effective charm-quark mass yields a smooth matching to the open-charm density of the 
hadronic phase using the hadron resonance gas model~\cite{Du:2015wha}.  The resulting effective charm-quark mass is approximately $1.46\,$GeV at $T=500\,$MeV and increases to about $1.76\,$GeV at $T_{\rm H}$. We emphasize that the effective mass is only used in the open-charm density for the sole purpose of obtaining the charm fugacity, to mimic the emergence of pre-hadronic states as the hadronization temperature is approached from above (the transport calculations are entirely based on the in-medium masses from the $T$-matrix analysis~\cite{Tang:2023tkm}), resulting in an effective mass that is noticeably smaller than the in-medium mass, cf.~Fig.~\ref{fig:mc_eff_T}, 
generating a larger open-charm density.
The contributions of (broad) resonance states are corroborated in earlier calculations of charm susceptibilities in the $T$-matrix approach~\cite{Liu:2021rjf}.
Even at temperatures as high as $500\,$MeV, the effective mass is still somewhat smaller than the in-medium mass, indicating the presence of resonance-like correlations as found in the complex pole analysis of the $T$-matrix with WLC constraints~\cite{Tang:2025ypa}.

\section{Charmonia at the LHC}
\label{sec:lhc}
We now turn to applications of our new transport framework to experiment by 
coupling it to a schematic fireball model for Pb--Pb collisions at $\sqrt{s_{\mathrm{NN}}}=5.02$~TeV at the LHC, focusing on the centrality and momentum dependence of charmonium production.
In Sec.~\ref{ssec:initial}, we specify the inputs to our approach, \ie, the charm/onium cross sections, including their transverse-momentum ($\pT$) dependence and cold-nuclear-matter modifications. In Sec.~\ref{ssec:equil-time} we discuss the time evolution of the charmonium equilibrium limits under the conditions of the fireball formed in Pb--Pb collisions. In Secs.~\ref{ssec:centrality} and \ref{ssec:pT} we show our results for the centrality and $\pT$-dependence of $J/\psi$, $\chi_c$ and $\psi(2S)$ production in comparison to experimental data and highlight the main differences to previous calculations.

The main observable is the nuclear modification factor, $\raa$, defined as the ratio of the charmonium yield in nucleus--nucleus (AA) collisions, $N_{\psi}^{\rm AA}$, to the one in proton--proton (pp) collisions scaled by the number, $N_{\text{coll}}$, of binary collisions for a given centrality selection,  
\begin{equation}
\raa^{\psi}(N_{\text{part}},\pT) =
\frac{N_{\psi}^{\rm AA}(N_{\text{part}},\pT)}
     {N_{\psi}^{\rm pp}(\pT)\, N_{\text{coll}}(N_{\text{part}})} \, ,
\label{eq:RAA_def}
\end{equation}
where $N_{\text{part}}$ is the number of participant nucleons characterizing the collision centrality.

\subsection{Initial conditions}
\label{ssec:initial}
%
\begin{figure}[t]
    \centering
    \includegraphics[width=0.47\textwidth]{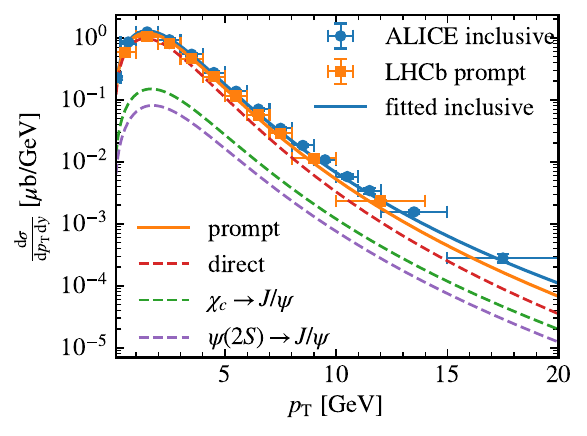}
    \caption{
    Inclusive (blue solid line) and prompt (orange solid line) $J/\psi$ transverse-momentum spectra in $pp$ collisions at $\sqrt{s}=5.02$~TeV.
    The prompt component is further decomposed into the direct contribution
    (red dashed line) and feeddown from $\chi_c$ (green dashed line) and
    $\psi(2S)$ (purple dashed line).
    The data are taken from ALICE~\cite{ALICE:2021qlw} and
    LHCb~\cite{LHCb:2019eaj}.
    }
    \label{fig:jpsi_pp}
\end{figure}
The differential $J/\psi$ production cross section in 5.02\,TeV $pp$ collisions has been measured as $\mathrm{d}\sigma_{J/\psi}/\mathrm{d}y = 5.64~\mu\mathrm{b}$~\cite{ALICE:2019pid} at mid-rapidity and
$\mathrm{d}\sigma_{J/\psi}/\mathrm{d}y = 3.93~\mu\mathrm{b}$~\cite{ALICE:2021qlw} at forward rapidity, $2.5<y<4$.
For excited states, we adopt the direct-production cross-section ratio in $pp$ collisions for $\psi(2S)$ to $J/\psi$ as
$N^{\rm pp}_{\psi(2S)}/N^{\rm pp}_{J/\psi} = 0.147$~\cite{ALICE:2021qlw} and
$N^{\rm pp}_{\chi_c}/N^{\rm pp}_{J/\psi} = 0.926$ for $\chi_c$~\cite{ParticleDataGroup:2024cfk}.

For the initial charmonium phase-space distribution, $f\!\left(\mathbf{x}, \mathbf{P}_{\Psi},t_0\right)$, we assume a factorization into three components: the spatial distribution, evaluated from a collision profile based on a Glauber model~\cite{Miller:2007ri}, the momentum spectrum, based on experimental data in 
$pp$ collisions~\cite{ALICE:2021qlw,LHCb:2021pyk,LHCb:2012af,LHCb:2012geo}, and the $\pT$-dependent shadowing estimated from ALICE $J/\psi$ data~\cite{ALICE:2015sru} in pPb collisions (discussed below).
%
%
We fit the inclusive $J/\psi$ $\pT$ spectrum measured at forward rapidity ($2.5<y<4$) in 5.02\,TeV $pp$ collisions~\cite{ALICE:2021qlw} with a power-law ansatz
\begin{equation}
\frac{\mathrm{d}N^{\rm pp}}{\mathrm{d}\pT^2} =
\frac{N}{\left[1+\left(\pT/A\right)^c\right]^n} \, ,
\label{eq:powerlaw}
\end{equation}
with parameters $N$=0.0599, $A$=3.81, $c$=2, and $n$=3.73;
at mid-rapidity, the corresponding fit to ALICE data~\cite{ALICE:2019pid} yields
$N$=0.043, $A$=4.48, $c$=2, and $n$=3.73.
The inclusive $J/\psi$ yield can be decomposed into direct production, prompt feeddown from excited states ($\chi_c$ and $\psi(2S)$), and non-prompt feeddown from bottom-hadron decays.
The non-prompt feeddown fraction is constrained by LHCb measurements at $\sqrt{s}=5.02~\mathrm{TeV}$~\cite{LHCb:2021pyk}, while the prompt feeddown fractions are taken from LHCb data at $\sqrt{s}=7~\mathrm{TeV}$~\cite{LHCb:2012af,LHCb:2012geo}.
Figure~\ref{fig:jpsi_pp} shows the resulting inclusive and prompt $J/\psi$ $\pT$ spectra in $pp$ collisions at $\sqrt{s}$=5.02\,TeV and forward rapidity, together with their decomposition into direct and feeddown contributions.
For the inclusive $\psi(2S)$ spectrum at forward rapidity in $pp$ collisions~\cite{ALICE:2021qlw}, the power-law fit yields
$N$=0.033, $A$=5.10, $c$=2, and $n$=3.70.
The non-prompt feeddown contribution to inclusive $\psi(2S)$ production is constrained using LHCb measurements at $\sqrt{s}=7~\mathrm{TeV}$~\cite{LHCb:2019eaj}.


\begin{figure}[t]
    \centering 
    \includegraphics[width=0.47\textwidth]{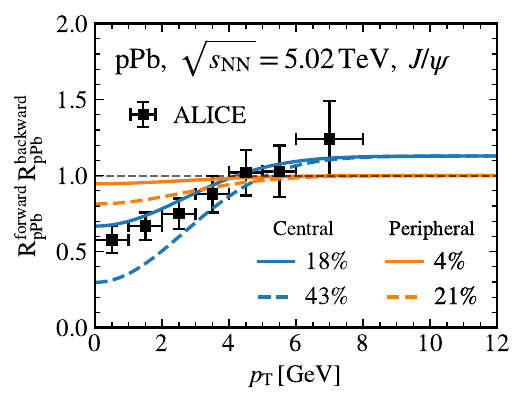}
    \hfill
    \includegraphics[width=0.47\textwidth]{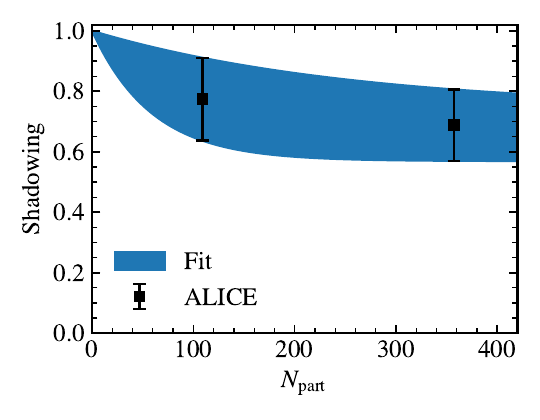}  
    \caption{
    Parameterizations of the nuclear-shadowing suppression of the $c \bar c$
    production cross section.
    The upper panel shows the $\pT$ dependence of $J/\psi$ production in $p$--Pb collisions compared to ALICE data~\cite{ALICE:2015sru},
    while the lower panel shows the charm cross section as a function of $N_{\mathrm{part}}$ in Pb--Pb collisions obtained from a fit to prompt $D^0$ $\raa$ data from ALICE~\cite{ALICE:2021rxa}.}
    \label{fig:shadow}
\end{figure}
The $c\bar{c}$ production cross section in $pp$ collisions is a key input for the
charmonium equilibrium limit and thus controls the regeneration yield.
For a given rapidity interval, the number of produced $c\bar{c}$ pairs is
given by $\dd N_{c\bar c}/\dd y = (\mathrm{d}\sigma_{c\bar c}/\mathrm{d}y)\,N_{\mathrm{coll}}/\sigma^{\rm inel}_{\rm pp}$, where $\sigma^{\rm inel}_{\rm pp}$ is the inelastic cross section in $pp$ collisions.
At mid-rapidity, the measured charm cross section amounts to $\mathrm{d}\sigma_{c\bar c}/\mathrm{d}y = 1.165 \pm 0.133~\mathrm{mb}$~\cite{ALICE:2021dhb}.
We extrapolate this to forward rapidity using the rapidity dependence of Ref.~\cite{Bierlich:2023ewv},
yielding $\mathrm{d}\sigma_{c\bar c}/\mathrm{d}y = 0.72 \pm 0.07~\mathrm{mb}$, where a $10\%$ uncertainty is assigned.
The initial spatial distribution of $c$-quarks is sampled from the Glauber model~\cite{Miller:2007ri}, while the transverse momenta ($\pT$) are from
Fixed-Order Next-to-Leading-Logarithm (FONLL) calculations~\cite{Cacciari:2012ny}, which we fit with the same ansatz as in Eq.~(\ref{eq:powerlaw})
resulting in the parameter values $N$=0.045, $A$=3.3, $c$=2.2, and $n$=2.7. The longitudinal momentum distribution is assumed to be thermal,
with periodic
boundary conditions imposed in the longitudinal direction of the fireball, such
that a quark exiting from one side re-enters from the opposite side.

Finally, we account for cold-nuclear-matter (CNM) effects which
modify both the magnitude and $\pT$ dependence of the charm(onium) production cross sections in the Pb--Pb environment.
The impact on the $c$-quark $\pT$ spectrum is implemented through a multiplicative modification factor adopted from
Ref.~\cite{Emelyanov:1997guf},
\begin{equation}
\begin{aligned}
S_c(\pT)
= &a_1\, e^{-b_1\,(\pT - d_1)}
+ c_1\\
+ &a_2\, e^{-b_2\,(\pT - d_2)}\,,
\end{aligned}
\end{equation}
where the parameters are given as
\[
\begin{aligned}
    a_1 &= -0.27,  & b_1 &= 0.17, & d_1 &= 1.00, \\
    a_2 &= -65.39, & b_2 &= 0.52, & d_2 &= -11.14, \\
    c_1 &= 1.02.   &     &        &
\end{aligned}
\]

The CNM effect for charmonium $\pT$ spectra is estimated using ALICE data~\cite{ALICE:2015sru}
on $J/\psi$ production in $p$-Pb collisions at forward and backward rapidities,
$2.5 < |y| < 4$.
We fit the product of the measured forward and backward nuclear modification factors,
$R_{p\mathrm{Pb}}$, which can be interpreted as the net effect of shadowing in Pb--Pb collisions, as shown in the upper panel of Fig.~\ref{fig:shadow}.

For the total charm cross section we estimate the shadowing from ALICE $D^0$ data in Pb-Pb (5.02\,TeV) by fitting a band to the available $\raa$ data~\cite{ALICE:2021rxa}
using a functional form as in previous work~\cite{Zhao:2008pp}: 
$S(\npart)= 1-a+a\exp(-N_{\rm part}/b)$, where $(a,b)=(0.26,257.3)$ and
$(0.43,58.8)$ for the upper and lower limits,
see lower panel in Fig.~\ref{fig:shadow}.
For 0--20\% central collisions the suppression amounts to 18--43\% of the integrated yield, and we assume a $\pT$ dependence that reproduces the forward-backward $R_{p\mathrm{Pb}}$ product.
For 40--90\% peripheral collisions, the shadowing effect is reduced
to 4--21\%, where, in contrast to central collisions, no anti-shadowing behavior is assumed at large $\pT$.
For other centrality classes considered below, \ie,  40--60\%, 30--50\% and 20-40\%, the shadowing ranges are 6--30\%, 9--36\% and 12--43\%, respectively. 

\subsection{Time evolution of $\Psi$ yields}
\label{ssec:equil-time}
\begin{figure}[t]
    \centering
\includegraphics[width=0.47\textwidth]{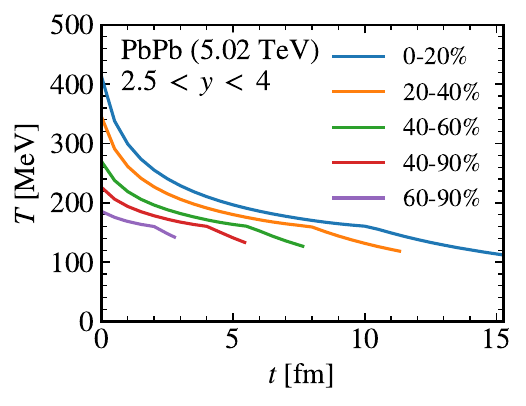}    

    \vspace{-0.8cm}  

\includegraphics[width=0.47\textwidth]{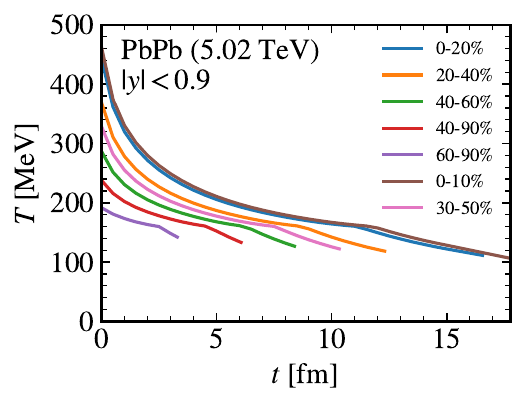}
\caption{Time evolution of the temperature of the expanding-fireball model at different centralities at forward (upper panel) and mid-rapidity (lower panel). At a ``hadronization'' temperature of $T_{\rm H}=170\,\text{MeV}$, the equation of state is switched from lattice QCD to a hadron resonance gas model (not visible in the plots). The kink in each curve occurs at the hadro-chemical freeze-out temperature of $T_{\rm ch}$=160~MeV, below which effective hadronic chemical potentials ensure the conservation of the observed light-hadron yields.}
    \label{fig:fireball}
\end{figure}

To enable a better understanding of how the charm-charmonium kinetics manifests itself under the conditions of heavy-ion reactions, and ultimately in the final charmonium yields, we first discuss the time dependence of their various components in the hot fireballs with the inputs specified in the previous section. For now, we employ a schematic (isotropic and isentropic) fireball model where the total entropy is fixed by the finally observed light-hadron abundances at a given centrality while the expansion is simulated by a time-dependent blast-wave description whose parameters are chosen to mimic hydrodynamic evolutions with endpoints (\ie, thermal freezeout) that are in the range of extractions from experimental $\pT$ spectra, cf.~also Ref.~\cite{Wu:2024gil}. The resulting temperature evolution is shown in Fig.~\ref{fig:fireball} at mid- and forward rapidities and for different centrality classes.  Clearly, future studies will have to improve on this framework.  

\begin{figure}[!t]
    \centering
    \includegraphics[width=0.47\textwidth]{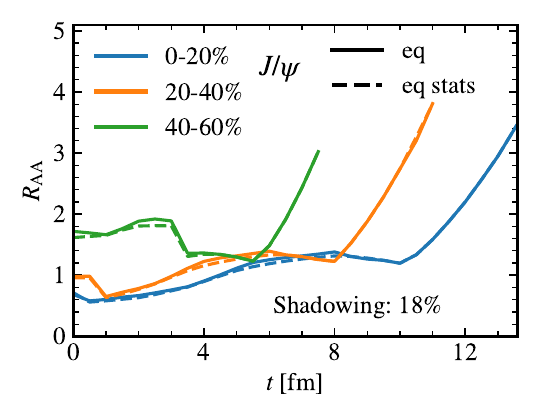}
     \caption{Equilibrium limit of $J/\psi$ as a function of time for different collision centralities (colors) assuming a shadowing of up to 18\% in central collisions.
     Dashed lines correspond to the equilibrium limit from the statistical model, while solid lines show the result from our transport approach in the charm-quark thermalization limit.
     }
    \label{fig:eq_stats}
\end{figure}
\begin{figure}[t]
    \centering
    \includegraphics[width=0.47\textwidth]{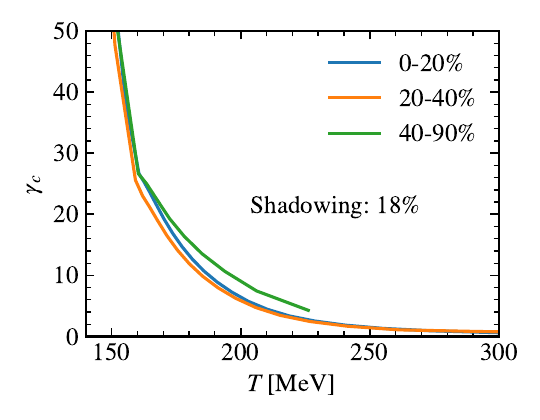}
    \caption{Charm-quark fugacity as a function of temperature in Pb-Pb($\sqrt{s_{\mathrm{NN}}}=5.02$~TeV) collisions at different centralities and forward rapidity.
    }
    \label{fig:fugacity}
\end{figure}

To begin with, we display in Fig.~\ref{fig:eq_stats} the $\raa$ for the $J/\psi$ equilibrium limit (without any feeddown) obtained from our coupled transport approach assuming fully thermalized charm quarks, and compare that to the results from the statistical model, as a function of time at different centralities (a similar comparison has also been reported in Ref.~\cite{Zhao:2025bzl}).
The two results are in good agreement and consistent with the analytical derivation presented above. The $\raa$ values are rather large, close to one or even higher, and two structures are discernible: a drop-off at relatively early times (except for central collisions), which is an effect of the correlation volume, and a late-time increase, which commences at the chemical freezeout temperature, $T_{\rm ch}$ in the hadron resonance gas equation of state (EoS).
We also note that the use of the lQCD EoS in our current calculations, compared to the quasiparticle approximation adopted in previous work~\cite{Wu:2024gil,Zhao:2007hh}, leads to a significant decrease of the equilibrium limits when the QGP approaches the hadronization temperature of $T_{\rm H}=170$\,MeV from above. The reason is the larger entropy density in the quasiparticle EoS, which converts into a smaller temperature at a given fireball time, resulting in larger fugacities, $\gamma_c$, thus increasing the charmonium equilibrium limits (which are proportional to $\gamma_c^2$). We illustrate the temperature dependence of the fugacity in our current calculation in Fig.~\ref{fig:fugacity}. 

\begin{figure*}[t]
    \centering
    \captionsetup{width=0.99\linewidth}
    \includegraphics[width=0.98\textwidth]{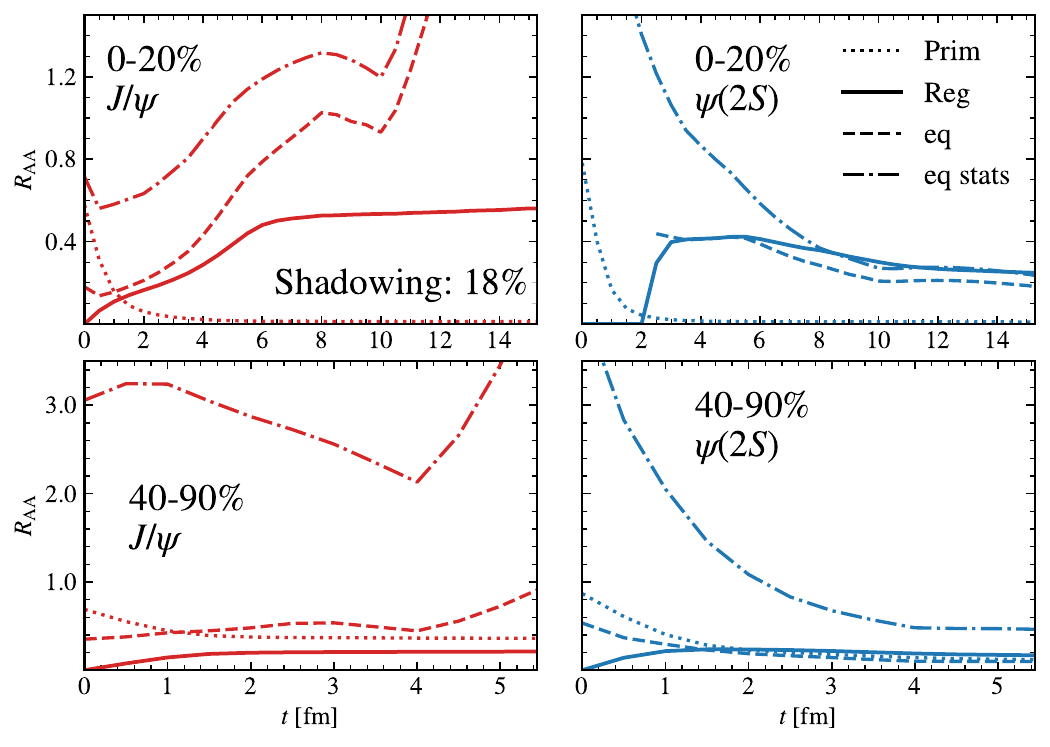}
    \caption{Time evolution of charmonium $\raa$ of 1S (left column) and 2S (right column)
    in central (top row) and peripheral (bottom row) Pb--Pb collisions at
    $\sqrt{s_{\mathrm{NN}}}=5.02\,\mathrm{TeV}$ at forward rapidity.
    The dash-dotted lines are the equilibrium limit from the statistical model,
    while the dashed lines are the equilibrium limit including off-equilibrium effects.
    Solid and dotted lines represent regenerated and primordial components,
    respectively.
    }
    \label{fig:time_evolution_Neq_static_statistic}
\end{figure*}

       
    

\begin{figure*}[!tbp]
    \centering

    

    \includegraphics[width=\textwidth]{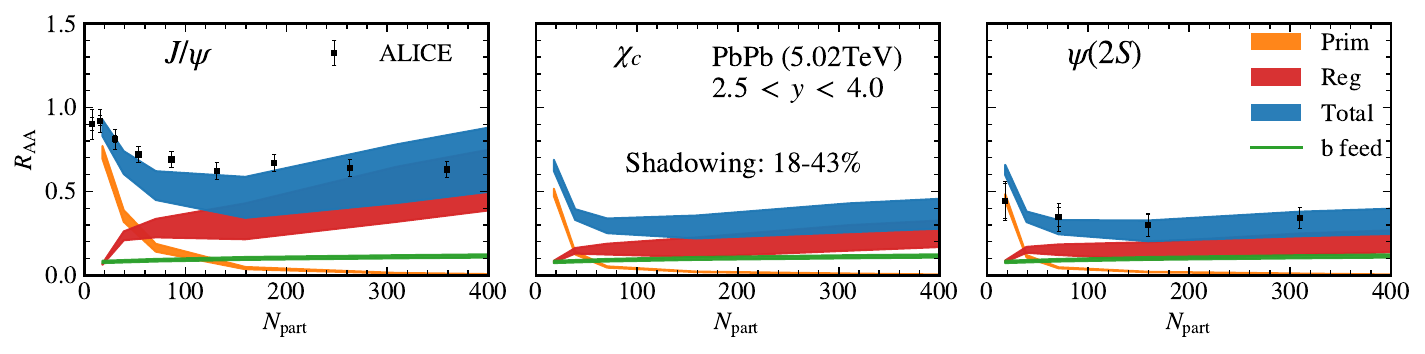}

    \vspace{-0.3cm}
    
    \includegraphics[width=\textwidth]{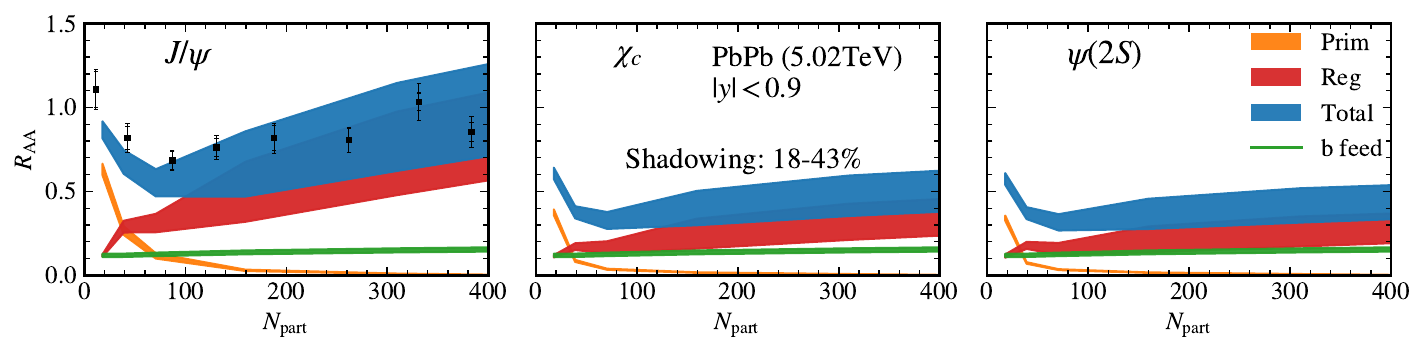}

    \caption{Centrality dependence of the inclusive nuclear modification factors for $J/\psi$ (left column), $\chi_c$ (mid column) and $\psi(2S)$ (right column) production in Pb--Pb (5.02 TeV) collisions at forward/backward (top row) and mid (bottom row) rapidity. The bands for the primordial (orange) and regenerated (red) components include uncertainties from nuclear shadowing on both open and hidden charm initial production up to a maximum of 18(43)\% in central collisions for the upper (lower) end of the bands, and for $B$ feeddown (green bands)  a 0--15\% uncertainty range. The totals (blue bands) are compared to ALICE data for $J/\psi$ and $\psi(2S)$~\cite{ALICE:2016flj,ALICE:2022jeh,ALICE:2023gco}. 
    }
    \label{fig:npart_bands}
\end{figure*}

With the equilibrium benchmark in place, we show in Fig.~\ref{fig:time_evolution_Neq_static_statistic} the time evolution of the primordial, regenerated and equilibrium limits of 
charmonium $\raa$'s in central and peripheral Pb--Pb ($\sqrt{s_{\mathrm{NN}}}=5.02$\,TeV) collisions at forward rapidity. Here, the equilibrium limit is obtained from the detailed balance relation, Eq.~\eqref{eq:detail_balance}, with time-dependent off-equilibrium charm-quark momentum distributions as follows from the Langevin simulations.
Compared to the statistical-equilibrium limit, it is substantially reduced, with the effect being more pronounced in peripheral collisions where the shorter fireball lifetime and lower temperatures limit $c$-quark thermalization, compared to central collisions. This reduction is understood to be a consequence of harder $c$-quark distributions, which have less favorable phase space for bound-state formation~\cite{Song:2012at} (it was implemented in our previous works through a thermal-relaxation time approximation~\cite{Grandchamp:2002wp}). 
In our current setup we do not account for open-charm diffusion in the hadronic phase; we therefore utilize the reduction of the equilibrium limit from the QGP phase at $T_H=170$~MeV 
to continuously match it to the hadronic phase and use the same rescaling factor throughout the hadronic medium evolution. The regeneration rate in the hadronic phase is then computed from the rescaled equilibrium limit and the hadronic reaction rate through the detailed-balance relation~\cite{Du:2015wha}.

In central collisions, both primordial $J/\psi$ and $\psi(2S)$ states experience strong suppression due to the large reaction rates in the strongly coupled QGP, and their primordial yields are essentially eliminated at early times. As a result, the final yields are dominated by regeneration. The regenerated $J/\psi$ gradually approaches its equilibrium limit during the QGP evolution. After $\sim 6\,{\rm fm}/c$, when the temperature has dropped below $\sim$190\,MeV and the reaction rate is below 100\,MeV, the regeneration yield of the $J/\psi$ levels off.
On the other hand, the $\psi(2S)$, due to its larger rates, reaches its equilibrium limit within $\sim 1$–$2\,{\rm fm}/c$ after regeneration sets in and stays rather close to it also at later stages.
In peripheral collisions, the timescales are much shorter, which implies much reduced regeneration while a significant portion of the primordially produced states survives.

\subsection{Centrality dependence}
\label{ssec:centrality}
We are now in position to extract the $\raa$ observables from our coupled charm-charmonium transport framework, starting with the centrality dependence of charmonium production at forward/backward and mid-rapidity, summarized in the upper and lower panels of Fig.~\ref{fig:npart_bands}, respectively.
For the $J/\psi$, the strong dissociation 
rate at high temperatures leads to a pronounced suppression in semi-central and 
central collisions, while the increasing charm-quark multiplicity toward 
central events enhances the regeneration contribution. As a result, the total 
$\raa$ develops a broad minimum around $N_{\rm part} \simeq 100-150$ followed 
by a rise toward most central collisions. This trend is more pronounced at mid-rapidity and is consistent with 
the picture that primordial charmonia are largely depleted in the hottest QGP 
regions, whereas subsequent recombination partially compensates for the loss. There is a slight tendency to underestimate the data at mid to peripheral centrality, where multiple pairs of charm quarks are produced but not in sufficient numbers to reach the grand canonical limit. Improved treatments of the canonical ensemble with correlation volume may be required here. 

For $\psi(2S)$, the suppression is significantly stronger due to its weaker binding and ensuing larger rates. The regeneration contribution is smaller compared to $J/\psi$, mostly reflecting the reduced statistical weight.
Consequently, after a steep initial suppression in peripheral collisions, the total $\raa$ exhibits a much weaker rise with centrality. The model 
reproduces the overall magnitude and slope of the ALICE measurements at forward rapidity fairly well.

    
   




\begin{figure*}[t]
    \centering
    \includegraphics[width=\textwidth]{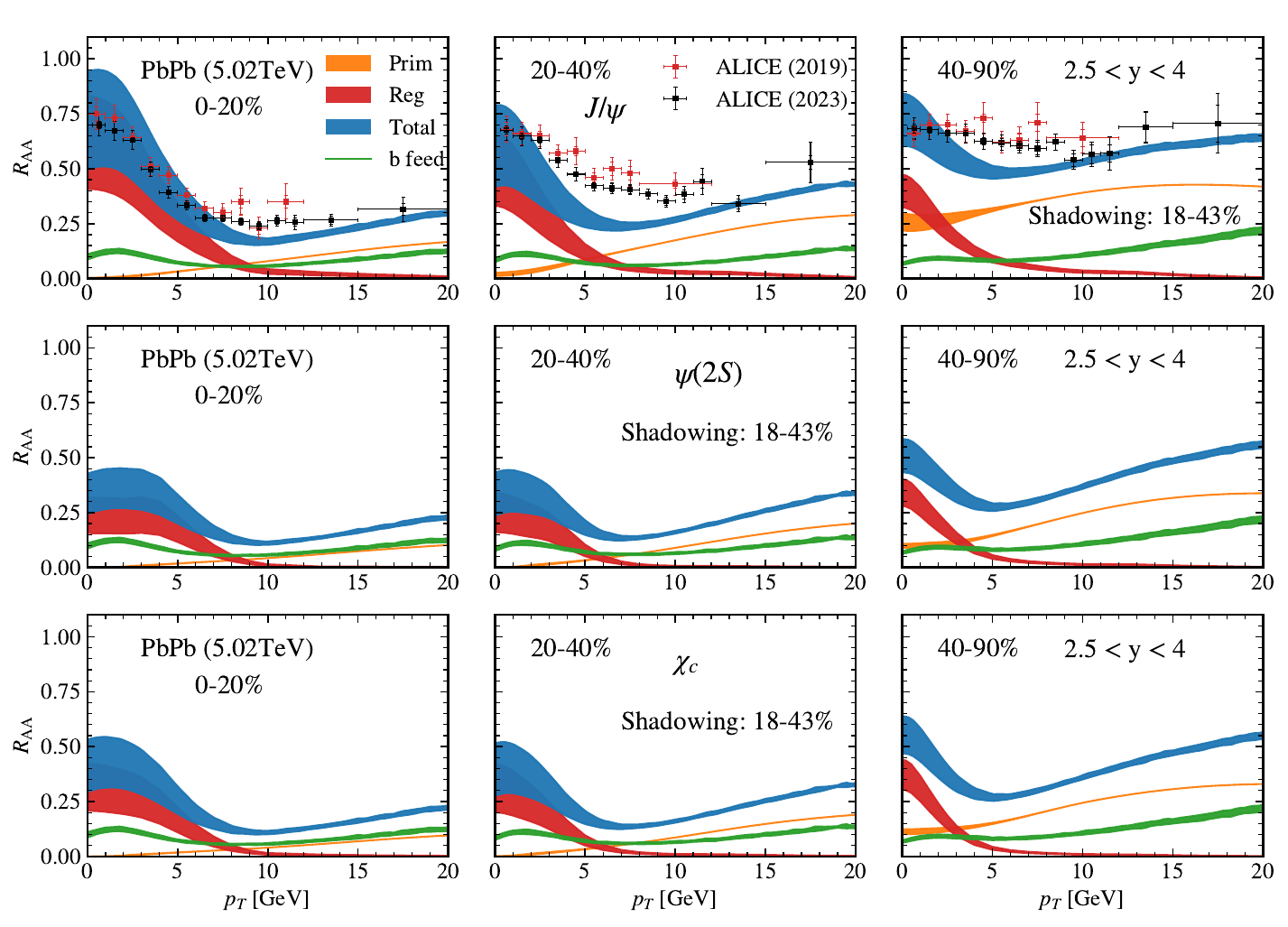}

    \caption{Transverse-momentum–dependent nuclear modification factor, $\raa$, of inclusive $J/\psi$ (top row), $\psi(2S)$ 
    (middle row) and $\chi_c$ (bottom row) production in Pb--Pb ($\sqrt{s_{\rm NN}}$=5.02\,TeV) collisions at forward/backward rapidity. The bands represent the primordial (orange), regenerated (red), $B$ feeddown (green) and total (blue) contributions, including uncertainties from nuclear shadowing. The calculations are compared to ALICE data~\cite{ALICE:2019lga,ALICE:2023gco}. 
    }
\label{fig:pt-for-y}
\end{figure*}

The $\chi_c$ results follow a pattern very similar to that of $\psi(2S)$: a strong primordial suppression and a rather moderate regeneration component with a slightly stronger increasing trend compared to the $\psi(2S)$ but still significantly less than the $J/\psi$. Since $\chi_c$ feeddown constitutes a sizable fraction of the inclusive $J/\psi$ yield in $pp$ collisions, 
its suppression also plays a role in shaping the observed $J/\psi$ spectra.


\begin{figure*}[htbp]
    \centering
    \includegraphics[width=\textwidth]{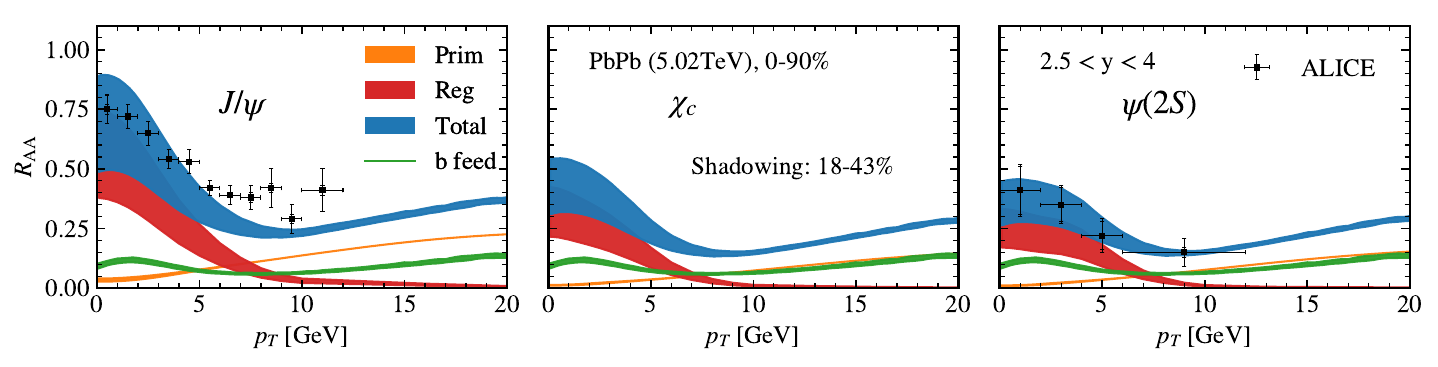}

    \caption{Same as Fig.~\ref{fig:pt-for-y} but for minimum-bias (0-90\%) Pb--Pb ($\sqrt{s_{\rm NN}}$=5.02\,TeV) collisions
    for $J/\psi$ (left), $\chi_c$ (middle) and $\psi(2S)$ (right) mesons.
    The calculations of $J/\psi$ and $\psi(2S)$ are compared to ALICE data~\cite{ALICE:2016flj,ALICE:2022jeh}.
    }
    \label{fig:pt-for-y-mb}
\end{figure*}

\subsection{Transverse-momentum spectra}
\label{ssec:pT}
\begin{figure*}[t]
    \centering
    \includegraphics[width=\textwidth]{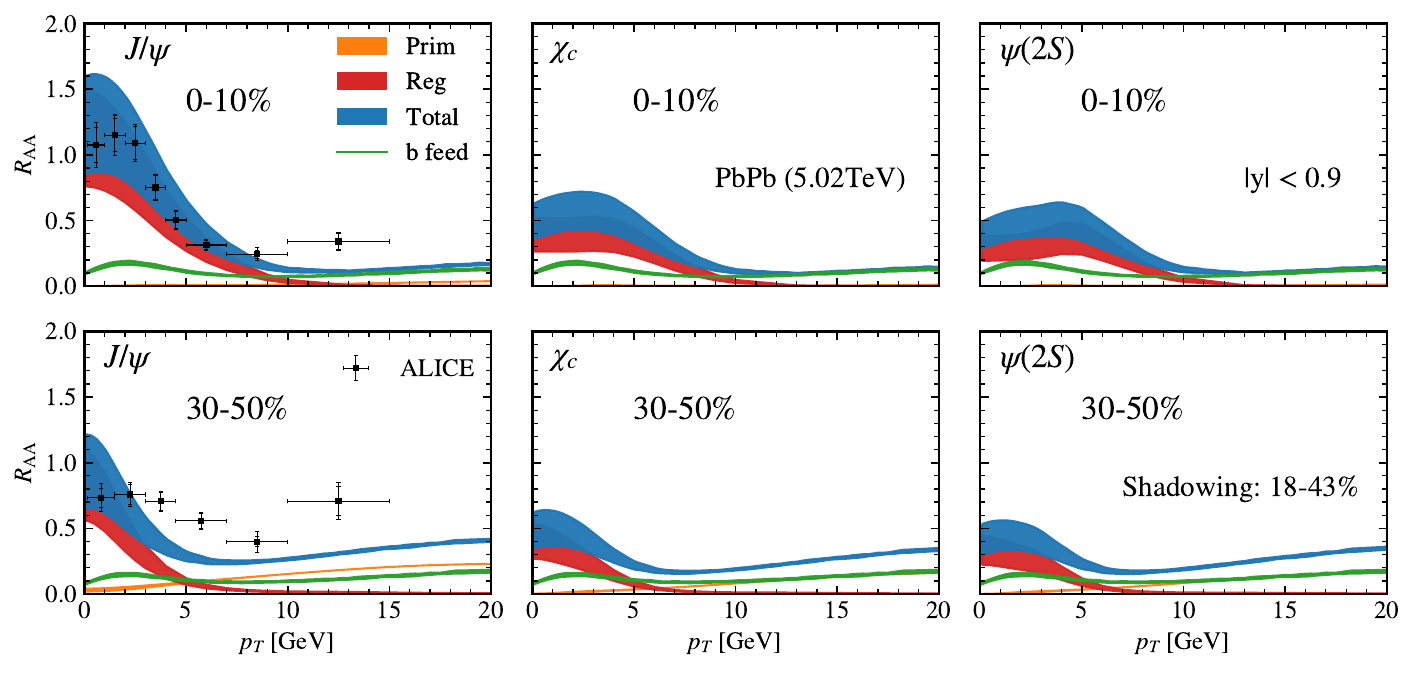}

    \caption{
    Transverse-momentum dependent $\raa$ at mid-rapidity for
    inclusive $J/\psi$ (left column), $\chi_c$ (mid column), and $\psi(2S)$
    (right column) production in Pb--Pb ($\sqrt{s_{\rm NN}}$=5.02~TeV) collisions, for
    0-10\% (top row) and 30-50\% (bottom row) centrality.
    Line identifications as in Fig.~\ref{fig:pt-for-y}. The $J/\psi$ calculations are compared to ALICE data~\cite{ALICE:2023gco}. 
    }
    \label{fig:pt_mid-y}
\end{figure*}

The interplay of (suppressed) primordial production and regeneration is expected to leave distinct signatures in the charmonium $\pT$-spectra~\cite{Zhao:2007hh}. In Figs.~\ref{fig:pt-for-y}, \ref{fig:pt-for-y-mb}, and \ref{fig:pt_mid-y} we summarize our results for the $\raa(\pT)$'s of $J/\psi$, $\psi(2S)$, and $\chi_c$ states at forward rapidity and mid-rapidity, respectively.
In addition to primordial suppression and regeneration, we estimate the nuclear modification of the $B$-feeddown contribution through Langevin simulations of $b$ quarks as described at the end of Sec.~\ref{sec:langevin}. 

For the $J/\psi$ the low-$\pT$ enhancement due to regeneration, followed by a transition to a primordial component, reproduces ALICE data fairly well. There is a significant underestimation of the data in the transition regime for peripheral and mid-central collisions, a feature that we also found in our previous calculations with a perturbative medium coupling. This may be remedied by the inclusion of space-momentum correlations (SMCs) between the diffusing $c$-quarks (where faster quarks are preferentially found in the outer parts of the fireball) which are not accounted for in our present treatment, but which largely resolved this problem in a hydrodynamic treatment in Ref.~\cite{He:2021zej}. On the other hand, the overestimate of the low-$\pT$ data in peripheral collisions is much less severe than in our previous calculations~\cite{Wu:2024gil} owing to the use of off-equilibrium $c$-quark spectra, as opposed to the blast-wave approximation in our previous work. The $\raa$'s of the excited states share largely similar features, although the low-$\pT$ enhancement is less pronounced since the regeneration is limited by the significantly smaller equilibrium limits at the $\raa$ level. The only data available for those are for the $\psi(2S)$ in minimum-bias collisions, cf.~the left panel in Fig.~\ref{fig:pt-for-y-mb}.  

A noteworthy yet somewhat subtle feature can be identified when comparing the low-$\pT$ flow bump due to regeneration of the different states in central collisions. For the $\psi(2S)$ the $\raa(\pT)$ is essentially flat over the first 4 GeV whereas for the $J/\psi$ it falls off almost immediately, by nearly 50\% at $\pT$=4\,GeV, and for the $\chi_c$ it is in between. This is likely a signature of the sequential regeneration mechanism first proposed in Ref.~\cite{Du:2015wha}, \ie,  since regeneration occurs at later times (lower temperatures) for less bound states, the latter will inherit harder spectra from the larger flow in the recombining charm and anticharm quarks.

For all three states, the suppression of primordial production at high $\pT$ is less pronounced than at low $\pT$. This is due to the reduced rates at high $\pT$ as well as Lorentz time dilation on the formation time effects during the early expansion of the charmonium wave packet. The escape effect, \ie, primordial charmonia leaving the medium, also leads to less suppression at high $\pT$, although this effect is less important than the former two.

We finally note the modulation of the $B$-feeddown contribution,
featuring a high-$\pT$ suppression accompanied by an accumulation at low $\pT$ that becomes more pronounced in more central collisions, although its role in the overall picture is not very relevant.



The results at mid-rapidity, displayed in Fig.~\ref{fig:pt_mid-y},
largely mirror the findings at forward $y$, although with some increase in the $\raa$ values at low $\pT$ due to the stronger regeneration at larger charm cross section and stronger suppression reflecting the longer fireball lifetime and higher initial temperatures.
For the 30-50\% centrality class, the same problem as for the mid-central forward-$y$ calculations is apparent, namely that the $J/\psi$ $\raa$ underestimates the data in the intermediate-$\pT$ region around 5\,GeV.
Again, a more realistic hydrodynamic medium and, in particular, SMCs are likely to improve on this.


\section{Conclusions}
\label{sec:conclusions}
In the present work we have developed a coupled transport approach where the diffusion of heavy quarks is directly coupled to the kinetics of quarkonia in a strongly coupled QGP. Recent calculations of charm and charmonium transport coefficients based on the same underlying microscopic interactions have been implemented into a coupled set of Langevin-Boltzmann equations. While the latter correspond to a semiclassical approximation, the transport coefficients follow from nonperturbative $T$-matrix interactions whose in-medium driving kernel (\ie, HQ potential) is constrained by state-of-the-art Wilson-line correlators computed in thermal lattice-QCD. In particular, we account for quantum effects in the sQGP driven by the large collision rates of the charm quarks and thermal-medium partons, which are encoded in the transport coefficients by pertinent off-shell spectral functions. The charm-quark transport coefficients have recently been deployed in open HF phenomenology at the LHC and were found to be compatible with $D$-meson observables without the use of any $K$-factors. An essential feature of our semiclassical simulations is that they recover the correct (thermal and chemical) equilibrium limit for charmonia for the case of thermalized charm quarks for an arbitrary number of pairs in the system, which is critical given the large reaction rates in the sQGP. The charmonium equilibrium limits have been constrained by lattice-QCD data for open-charm susceptibilities, and an off-shell extension of the regeneration rate with in-medium spectral functions has been constructed that maintains detailed balance with earlier calculated dissociation rates. 
In this way, the impact of the gradual thermalization in the open-charm sector on charmonium kinetics can be controlled, thus replacing earlier employed relaxation-time approximations. The general feature of a suppression of regeneration due to incomplete $c$-quark thermalization persists, in semi-quantitative agreement with the relaxation time approximation but without free parameters.

We have conducted a preliminary application to charmonium data from Pb--Pb collisions at the LHC using a schematic fireball model. Compared to our previous work, the much larger rates substantially accelerate the suppression of the primordial component, especially at low $\pT$, essentially wiping it out in (semi-) central collisions up to about $\pT\simeq 5(2)$~GeV. At the same time, the regenerated yields reach close to the ``equilibrium limit" with incomplete $c$-quark thermalization.   
The latter is particularly relevant also in peripheral collisions, where the resulting, much flatter $\pT$ spectra remedy an overshoot at low $\pT$ that was found when using a blast-wave approximation.
Overall, the agreement with the $J/\psi$ data is fair, especially given that there was no tuning of transport parameters, \ie, the charmonium reaction rates and equilibrium limits, as well as the $c$-quark diffusion. The calculated yields tend to have a somewhat stronger increase with centrality than the data suggest, and a deficit at intermediate $\pT$ for semi-central collisions. The use of a more realistic viscous hydrodynamic medium evolution and the inclusion of space-momentum correlations are candidates to improve on that.


Several future developments, especially in the phenomenological implementation, are in order.
A first step is to embed the transport dynamics into hydrodynamic simulations of the space-time evolution of the bulk medium, and to incorporate the aforementioned space-momentum correlations between the medium and (fast-moving) charm quarks.
While quantum transport for multiple pairs remains very challenging, the current use of formation time effects in the evolution of the individual primordial wave packets warrants a more rigorous quantum treatment. Furthermore, simulations of the coupled dynamics of $c\bar{c}$ pairs as microscopic degrees of freedom, rather than treating charm quarks and charmonia as independent particles, should be further investigated, see, \eg, Refs.~\cite{Young:2008he,Oei:2024pva,Daddi-Hammou:2025hdz}, possibly to serve as mutual benchmarks.

\section*{Acknowledgments}
We thank X.~Du and J.~Stachel for helpful discussions and comments.
This work is supported by the U.S. National Science Foundation under grant nos.\,PHY-2209335 and 2514775, and by the Department of Energy via the Topical Collaboration in Nuclear Theory on \textit{Heavy-Flavor Theory (HEFTY) for QCD Matter} under award no.\,DE-SC0023547.

\bibliography{refcnew}

\end{document}